\documentclass[apj,twocolappendix, numberedappendix]{emulateapj}
\usepackage[colorlinks,citecolor=blue]{hyperref}
\usepackage{textcomp}
\usepackage{amsmath}
\usepackage{graphicx}
\usepackage{subfigure}
\usepackage{color}
\usepackage{verbatim}
\usepackage{bm}
\usepackage{hyperref}

\def \msun {M_{\odot}}

\shortauthors{F. Yuan et al. }
\begin{document}

\title{Active Galactic Nuclei Feedback in an Elliptical Galaxy with the most updated AGN physics (I): low-angular momentum case}
\author{Feng Yuan$^{1}$, DooSoo Yoon$^{1}$, Ya-Ping Li$^{1}$,  Zhao-Ming Gan$^{1}$,  Luis C. Ho$^{2,3}$, and Fulai Guo$^1$}
\affil{$^{1}$Key Laboratory for Research in Galaxies and Cosmology, Shanghai Astronomical Observatory, Chinese Academy of Sciences, 80 Nandan Road, Shanghai 200030, China; fyuan,yoon,liyp,zmgan,fulai@shao.ac.cn}
\affil{$^2$Kavli Institute for Astronomy and Astrophysics, Peking University, Beijing 100871, China; lho.pku@gmail.com}
\affil{$^3$Department of Astronomy, School of Physics, Peking University, Beijing 100871, China}
\shortauthors{F. Yuan et al.}


\begin{abstract}
We investigate the effects of AGN feedback on the cosmological evolution of an isolated elliptical
galaxy by performing two-dimensional high-resolution hydrodynamical numerical simulations. The
inner boundary of the simulation is chosen so that the Bondi radius is resolved.  Compared to
previous works, the two accretion modes, namely hot and cold, which correspond to different
accretion rates and have different radiation and wind outputs, are  carefully discriminated
and  the feedback effects by radiation and wind in each mode are taken into account.  The most
updated AGN physics, including the descriptions of radiation and wind from the hot accretion
flows and wind from cold accretion disks, are adopted. Physical processes like star formation,
Type Ia and Type II supernovae are taken into account.  We study the AGN light curve, typical AGN
lifetime, growth of the black hole mass, AGN duty-cycle, star formation, and the X-ray surface
brightness of the galaxy. We compare our simulation results with observations and find general
consistency. Comparisons with previous simulation works find significant differences, indicating
the importance of  AGN physics.  The respective roles of radiation and wind feedbacks  are examined
and it is found that they are different for different problems of interest such as AGN luminosity
and star formation. We find that it is hard to neglect any of them, so  we suggest to use the
names of ``cold feedback mode'' and ``hot feedback mode''  to replace the currently used  ones.
\end{abstract} 

\section{Introduction}

There are growing evidences for the co-evolution between the central supermassive black holes
and their host galaxies \citep{Kormendy:13}, including the strong correlation between the mass
of the black hole and the luminosity, stellar velocity dispersion, or the stellar mass in the
galaxy spheroid \citep{Magorrian:98, Tremaine:02, Haring:04, Gultekin:09}. It is generally
believed that AGN feedback plays an important role in the evolution of galaxies and results in
the co-evolution between black hole growth and galaxy evolution \citep{Fabian:12, Kormendy:13}.

Over the past decade many theoretical studies aiming at AGN feedback in galaxy formation and
evolution have been performed, especially using the approach of hydrodynamical numerical simulation
 \citep[e.g.][]{DiMatteo:05, Springel:05, Sijacki:07, Ciotti:07, Booth:09, Ciotti:10,
Ciotti:17, Ostriker:10, Debuhr:11, Hirschmann:14, Gan:14, Eisenreich:17, Weinberger:17a}. The
most serious challenge of these studies is the large dynamical range: a proper simulation
involving both the large-scale structure of the universe and the black hole accretion scales
would require a spatial range of over ten orders of magnitude, from the black  hole radius
($R_s$) of $\sim 10^{-5}\,{\rm pc}$ for a $10^8\msun$ mass black hole, to a scale of hundreds
of Mpc. This is still technically infeasible even with modern supercomputers. Consequently, if
the study focuses mainly on large scales, such as the galaxy or larger scales, as it is the case
for most AGN feedback works so far, since the scale of black hole accretion cannot be resolved,
so-called sub-grid models have to be used. This ``sub-grid'' model means that some simplifications
and approximations are used when describing the black hole accretion and its outputs.

The first issue the ``sub-grid'' model needs to deal with is the determination of the mass
accretion rate of the central black hole. The exact determination of the mass accretion rate
is crucial for the study of AGN feedback since it determines the strength of the AGN outputs
such as radiation, jet, and wind. Most works adopt the Bondi solution \citep{Bondi:52,
Frank:02} to calculate the mass accretion rate (e.g., \citealt{Springel:05, Booth:09,
McCarthy:10, McCarthy:17, Choi:12, Choi:15, LeBrun:14, Schaye:15}; see recent review by
\citealt{Negri:17}). But this approach suffers from the following problems. First, the
accretion gas is usually turbulent and inhomogeneous \citep[e.g,][]{Gaspari:13}. Second,
the Bondi solution neglects  the angular momentum of the gas \citep{Hopkins:11, Tremmel:17}.
Perhaps the most serious problem is that, since the Bondi radius  cannot be resolved, some form
of extrapolations to the density and temperature have to be used. This is in practice achieved
by introducing a constant ``accretion rate boost parameter $\alpha$''  with a typical value of
100 \citep[e.g.,][]{Springel:05}. The underlying idea is that at large radii, the density may be
underestimated while the temperature overestimated compared to the values at the Bondi radius,
so the accretion rate will be underestimated. Obviously, the value of $\alpha$ has very large
uncertainties. In fact, different works choose quite different values, from $\alpha\sim 1$
to $\sim 300$ \citep[e.g.,][]{Booth:09}. Most recently, taking the AGN feedback in an isolated
galaxy as an example, \citet{Negri:17} carefully studied how well the Bondi approximation
 describes the mass accretion rate. They have performed many runs with different resolutions
using the Bondi approximation, and compared these results with a high-resolution  run which
can well resolve the Bondi radius thus precisely determining the accretion rate. They find that
approximated Bondi formalism can lead to both over- and underestimations of the black hole growth,
depending on the resolution and how the variables entering into the formalism are calculated. So
how to determine the mass accretion rate of the central AGN is still an unsolved problem.

In this paper, we focus on discussing another aspect of the ``sub-grid'' model. That is, for
a given mass accretion rate, what will be the exact outputs, i.e., radiation, wind, and jet,
from the AGN? In many large-scale cosmological simulation works, once they estimate the mass
accretion rate, they estimate the total energy output of the accretion flow and then simply
assume some fraction of these energy is converted into the thermal energy of the surrounding ISM
\citep[e.g.,][]{Booth:09},  except in some recent works in which the importance of mechanical
feedback by wind has  been recognized and taken into account \citep[e.g.,][]{Choi:15,
Weinberger:17a, Weinberger:17b, Gaspari:17}.  In reality, however, the output of black hole
accretion is much more complicated. As we will discuss in detail in \S\ref{sec:agnphysics},
there are two kinds of accretion modes, namely cold and hot, and in each mode the three types of
outputs are quite different.  In the best cases,  the two different accretion (feedback) modes
are discriminated, but the most updated accretion physics describing their outputs in each mode
has not been properly taken into account.  This is especially the case for the hot accretion
mode, in which case much progress has been made  in  recent several years. For example, when
calculating the Compton heating, usually the same Compton temperature value is used for both
accretion modes. This is not correct because the typical spectrum emitted in the two modes is
different (\citealt{Xie:17}; see also \S\ref{sec:agnphysics} below). The properties of wind in
the two modes are also often  described by the same parameterized way, while in recent couple of
years the properties of wind launched from the hot accretion flow have been intensively studied
and well understood, which is found to be very different from the wind from cold accretion disks
(see \S\ref{sec:agnphysics} below for details).

By performing two-dimensional hydrodynamical numerical simulations, we in this paper study the
evolution of an isolated elliptical galaxy.  The inner boundary is chosen to resolve the Bondi
radius so that the mass accretion rate can be calculated precisely. We consider two accretion
modes, in each mode we consider both the radiation and wind.  Jets are neglected as we will
explain later in this paper. Heating and cooling, star formation, Type Ia and Type II supernovae
are included in the simulation. These features are the same as those of some previous works
\citep[e.g.,][]{Novak:11, Ciotti:12, Gan:14, Negri:17, Ciotti:17}.  However, different from
these previous works, in this paper we will adopt the most recent progresses in AGN physics
which give correct descriptions of radiation and wind. As we will see in \S\ref{sec:agnphysics},
these progresses are especially in the regime of hot accretion flows but also in the description
of wind from cold accretion disk.

The paper is organized as follows. In \S\ref{sec:agnphysics}, we present the most updated AGN
physics we use in the simulation. The other physics of the model and the details of the  model
setup are described in \S\ref{sec:Model}.  The results of the simulation, namely the effects
of the AGN feedback in the galaxy evolution, are described in detail in \S\ref{sec:Results}.
We will also compare our results with those obtained in  previous works to investigate the
effects of the  AGN physics. Attention will be also paid on the respective roles of radiative
and mechanical feedback. The last section is devoted to a  summary and conclusion. As the first
paper of a series in this project, in this work we assume the specific angular momentum of the
gas in the galaxy is low.  In a following paper \citep{Yoon:18}, we will investigate  the effect
of AGN feedback on the evolution of elliptical galaxies with a high specific angular momentum.

\section{The AGN physics}
\label{sec:agnphysics}

As in previous works, the feedback effects are considered by injecting radiation and wind launched
by the AGN in the innermost grids of the simulation domain. The subsequent interaction of radiation
and wind with the ISM in the galaxy is self-consistently calculated. In this section, we describe
in detail how the properties of radiation and wind are calculated for a given accretion rate. This
AGN physics is also often called  ``feedback physics'' or ``sub-grid physics'' in literature. We
emphasize that the most recent progresses  in the field of AGN study will be taken into account.

\subsection{The accretion rate and two accretion modes}

Since our simulation can  resolve the Bondi radius, we can calculate the accretion rate at that
radius directly.  Note that we should not put an additional constraint of ``Eddington limit'' to
the mass accretion rate as in some papers. The Eddington limit only applies for a pure spherical
accretion which is almost always an oversimplification when we consider accretion flow close
to the black hole.  Once the accretion flow has some angular momentum, the accretion rate can
be of any value.

Then our question is, for a given accretion rate, what will be the output of AGN, i.e., what
are the properties of emitted radiation and wind?

After several decades' efforts, tremendous progresses have been made in the theory of black
hole accretion \citep[see reviews by][]{Pringle:81, Frank:02, Blaes:14, Yuan:14}. We know that
according to the temperature of the accretion flow, we have two series of solutions, i.e., cold
(e.g., the standard thin disk; \citet{Shakura:73}) and hot (e.g., the advection-dominated accretion
flow; \citet{Narayan:95}). Mathematically, these two series of solutions have a large overlap
in terms of the corresponding accretion rate. In other words, for a given accretion rate, both
the cold and hot accretion solutions are available for some  accretion rates \citep{Narayan:95,
Yuan:14}. Fortunately, such a theoretical degeneracy is broken by our observations of black hole
X-ray binaries. Black hole X-ray binaries come in two main ``states'' -- soft and hard states
-- which are described by the standard thin disk and the hot accretion flow, respectively.
During the decay phase of an outburst of a black hole X-ray binary, the luminosity decreases
with time. It is interesting to note that  the source always transits from the soft to the hard
state once the luminosity passes through a critical luminosity
\begin{equation}
  L_{\rm c}\approx 2\% L_{\rm Edd}
  \label{criticall}
\end{equation}
\citep{McClintock:06}. In other words,  we never observe a soft state with $L_{\rm BH} \la
2\%L_{\rm Edd}$\footnote{This is correct only to zeroth order. Here we do not discuss the
complication due to ``hysteresis'' in the hard-to-soft state transition.}.  So this ``critical''
luminosity is a boundary set by nature between the cold and hot accretion flows.

We have the almost same situation in the case of AGNs since the accretion physics does not
depend on the mass of the black hole. For AGNs, we now know that luminous AGNs such as quasars
correspond to the soft state, and are described by the cold accretion flow. In this accretion
mode, accretion flow produces radiation and wind but not jet\footnote{The absence of jet in
the cold accretion mode is based on the observations of the soft state of black hole X-ray
binaries. It is widely believed that a standard thin disk is operating in this state, but no
jets are observed. However, we do observe strong radio emission from the radio-loud quasar in
which we also believe a cold accretion disk is operating. It is still an open question how
to understand radio-loud quasar.}. Low-luminosity AGNs correspond to the hard state and are
described by hot accretion flows \citep{Ho:08, Yuan:14}. In this accretion mode,  accretion
flow produces three kinds of output, namely radiation, wind, and jet \citep{Yuan:14}.

The output from the two modes of accretion is quite different and this obviously will determine
the different feedback effects.  In the literature of AGN feedback, the corresponding feedback
are called quasar (or radiative) and radio (or kinetic, or maintenance, or jet) modes,
respectively. The names of the two feedback mode are not only diverse, but also confusing.
They were invented perhaps to emphasize the dominance of some forms of AGN output in an accretion
mode, for example the radiation in the cold accretion mode and the jet  in the hot accretion
mode. However, as we can see from \S\ref{subsec:comparison}, it is not obvious why we can
neglect the other forms of output in a given accretion mode, e.g., the wind output in the cold
accretion mode and the  radiation output in the hot accretion mode, at least when we compare
the energy and momentum fluxes of various AGN outputs (wind and radiation). It is even on the
opposite with what we think if we consider the feedback effects. For example, in the quasar mode,
at least in terms of controlling the black hole growth, wind feedback is much more important
than radiation (\S\ref{subsec:BHgrowth})\footnote{In the present work,  the role of dust is
not considered. However, the inclusion of dust in the simulation can significantly increase
the coupling between radiation and gas, thus the potential role of radiation in AGN feedback,
including in controlling the black hole growth, will be significantly enhanced\citep{Novak:12,
Ishibashi:15, Bieri:17,Costa:17}.}.  While in the radio mode, when the accretion rate is low,
radiative feedback seems to be at least equally important with wind in controlling the accretion
rate of the black hole (\S\ref{subsec:lightcurve}).  Therefore, in this paper, we suggest to
simply follow the names of black hole accretion mode and replace these complicated feedback
names by ``cold feedback mode'' and ``hot feedback mode'', representing the feedback occurred
in the two accretion modes respectively.

The  mass accretion rate corresponding to the critical luminosity $L_{\rm c}$ is:
\begin{equation}
\dot{M}_{\rm c}\approx \frac{L_{\rm c}}{\epsilon_{\rm EM,cold}c^2}.
\label{criticalrate}
\end{equation}
Here $\epsilon_{\rm EM,cold}$ is the radiative efficiency of a cold accretion disk. Note that
this accretion rate is the rate at the innermost region of the accretion disk where most of the
radiation comes from. Since winds exist in a cold disk, the accretion rate at the Bondi radius
must be larger than this rate.  For our description of wind launched from a cold disk (refer to
\S\ref{subsubsec:colddiskwind}),  the ratio of accretion rate close to the black hole and the wind
mass flux is a weak function of accretion rate. For typical parameters of our model, the ratio is
close to unity, i.e.,  $\dot{M}_{\rm BH}\approx \dot{M}_{\rm W}$. So the  accretion rate at the
innermost radius of the simulation domain ($r_{\rm in}$)  is only 2 times of that close to the
black hole horizon.  Therefore,  in our simulation, we compare the accretion rate at $r_{\rm in}$
(calculated by Eq. \ref{mdotbondi}) with $\dot{M}_{\rm c}$  in Eq. (\ref{criticalrate}) to decide
which accretion mode the accretion flow should choose. In the next two subsections, we describe
the output from the cold and hot accretion modes.  In this work, we neglect the effect of jet
because we assume that for the feedback study of a single galaxy as in the present work, the jet
may simply pierce through the galaxy and has negligible interaction with the galaxy since it is
well collimated, although it should be important for the evolution of large scale structure such
as galaxy clusters. But we note that there are debates on this topic (e.g., see \citet{Gaibler:12}
for a different opinion) so it is necessary to examine this assumption in the future works.

\subsection{Cold accretion (feedback) mode}
\label{subsec:coldmode}

\subsubsection{The accretion within the Bondi radius}

Our simulation can resolve the Bondi radius, which is the outer boundary of the accretion
flow. But within Bondi radius, the accretion onto the black hole still cannot be resolved and must
be treated as ``sub-grid'' physics. The dynamics of accretion gas within Bondi radius is likely
complicated and is a good project of future research. In this work, we assume the following
simple scenario.  Since the specific angular momentum of the gas is low, the gas will first freely
fall until a small accretion disk is formed with the size of the circularization radius. Note
that the disk should gradually grow with time because of the angular momentum transport of the
accretion flow\citep{Bu:14}. Wind will be launched from the disk and will affect the black hole
accretion rate.  In the following we estimate the black hole accretion rate according to the
above scenario. The approach is largely adopted from \citet{Ciotti:12} but with some modifications.

The mass accretion rate at the innermost radial grid ($\dot{M}(r_{\rm in})$) of our simulation
is calculated by Eq. (\ref{mdotbondi}). The infall timescale from $r_{\rm in}$  is calculated by
\begin{equation}
    \tau_{\rm ff} = \frac{\rm r_{in}}{v_{\rm ff}}, ~~~
    v_{\rm ff} \equiv \left( \frac{2\,G M_{\rm BH}}{r_{\rm in}} \right)^{1/2} ,
\end{equation}
where $\tau_{\rm ff}$ and $\,v_{\rm ff}$ are free-fall time scale and velocity at the inner-most
grid $r_{\rm in}$. The effective accretion rate at which gas feeds the small accretion disk is then obtained from
\begin{equation}
    \frac{d\,\dot{M}^{\rm eff}}{d\,t} = \frac{\dot{M}(r_{\rm in}) - \dot{M}^{\rm eff}}{\tau_{\rm ff}}.
\end{equation}
This equation implies that when $\dot{M}(r_{\rm in})$ drops to zero the small accretion disk
experiences a fueling declining exponentially with time.  Given our fiducial numerical set-up
(i.e. $R_{\rm in}$=2.5 pc and $M_{\rm BH,init}\sim 10^{9}\,M_{\odot}$), the free-fall time is
$\tau_{\rm ff}\sim 10^{3}$ yr.

Once the gas reaches  the circularization radius, $R_{\rm cir}$, it forms an accretion disk and
fuels the black hole with the accretion (viscous) timescale. With the computed total mass of
the gas in the disk, $M_{\rm dg}$, the mass inflow rate at $R_{\rm cir}$ in the disk can be estimated as
\begin{equation}\label{eq:disk}
    \dot{M}_{\rm d,inflow} = \frac{M_{\rm dg}}{\tau_{\rm vis}},
\end{equation}
where $\tau_{\rm  vis}$ is the instantaneous viscous timescale at $R_{\rm cir}$, which is  described in \citet{Kato:08},
\begin{equation}
    \tau_{\rm vis}\approx 1.2\times 10^{6}\, {\rm yr}
        \left(\frac{\alpha}{0.1}\right)^{-1} \left( \frac{R_{\rm cir}}{100\,r_{\rm s}}\right)^{7/2}
        \left( \frac{M_{\rm BH}}{10^{9}\,M_{\odot}}\right),
        \label{eq:viscous}
\end{equation}
where we set the viscosity parameter to $\alpha=0.1$. We note that in the simulation, we set
$R_{\rm cir}$ as a free parameter and adopt the value of $R_{\rm cir} = 100\, r_{\rm s}$,
where $r_{\rm s} \equiv 2G M_{\rm BH}/{c^{2}}$ is the Schwarzschild radius. In this way, the
growth of $R_{\rm cir}$ with time is absorbed in this parameter.

While the gas approaches the central black hole via the accretion disk, a fraction of the gas
is ejected as a form of wind. So the final black hole mass accretion rate can be obtained from
\begin{equation}
    \dot{M}_{\rm BH} = \dot{M}_{\rm d,inflow} - \dot{M}_{\rm W,C},
    \label{eq:BHaccretionrate}
\end{equation}
where $\dot{M}_{\rm W,C}$ is the wind mass flux at the cold accretion disk(see eq.~(\ref{eq:coldwindmass})). We will discuss the
wind model in \S\ref{subsubsec:colddiskwind}.

The total amount of disk gas mass, $M_{\rm dg}$, evolves as
\begin{equation}\label{eq:dg}
    \frac{d\,M_{\rm dg}}{d\,t} = \dot{M}^{\rm eff} - \dot{M}_{\rm BH} - \dot{M}_{\rm W,C}.
\end{equation}
Combining  eqs.(\ref{eq:disk})-(\ref{eq:dg}), we can estimate the black hole accretion rate
$\dot{M}_{\rm BH}$.  We note that if the disk is in hot accretion mode, the viscosity time
scale is many orders of magnitudes shorter than that in the cold mode. So both $\tau_{\rm ff}$
and $\tau_{\rm vis}$ are very small. Hence, we simply ignore the above-mentioned time lag effects.

\subsubsection{Wind}
\label{subsubsec:colddiskwind}
Cold accretion disk can produce strong winds, which have been widely observed in luminous
AGNs \citep[e.g.,][]{Crenshaw:03, Tombesi:10, Tombesi:14, King:15, Liu:15} and black hole
X-ray binaries (e.g., \citealt{Neilsen:12, Homan:16}, see review by \citealt{DiazTrigo:16}).
Three mechanisms have been proposed to explain the production of wind, namely
thermal, radiation, and magnetic ones. Debates still exist for which one is the dominant mechanism
in various conditions. In principle by performing numerical simulations we could combine all
these three mechanisms to calculate the wind properties. Unfortunately, it is still infeasible
due to the technical difficulties of globally simulating a thin disk.

Therefore we will obtain the properties of wind to be used in the current work from
observations. Some parameters of wind have been measured, such as the velocity, although the value
is diverse.   Recently, \citet{Gofford:15} have analysed a sample of 51 {\it Suzaku}-observed
AGNs and presented the properties of wind as a function of the bolometric luminosity of AGNs.
Their results indicate that more luminous AGN tends to harbour faster and more energetic winds.
In this paper, we adopt the fitted formulas presented in that work to describe the wind. The mass,
momentum and energy fluxes of the winds are described by\footnote{Note that in the Table 4 of
\citet{Gofford:15}, the value of the power index in the mass flux fitting formula is 0.9. The
value of 0.85 we adopt here is within the truncation error of their value and is more consistent
with the fitted line in their plots.},
\begin{equation}\label{eq:coldwindmass}
    \dot{M}_{\rm W,C} = 0.28\,  \,\left( \frac{L_{\rm BH}}{10^{45}\,\rm erg\,s^{-1}} \right)^{0.85}M_{\odot}\,{\rm yr^{-1}},
\end{equation}
\begin{equation}\label{eq:coldwindmom}
    \dot{P}_{\rm W,C} = \dot{M}_{\rm W,C}\,v_{\rm W,C},
\end{equation}
\begin{equation}\label{eq:coldwindenergy}
    \dot{E}_{\rm W,C} = \frac{1}{2}\, \dot{P}_{\rm W,C}\, v_{\rm W,C}.
\end{equation}
Here $L_{\rm BH}$ is the bolometric luminosity of the AGN. The velocity of wind is described by,
\begin{equation}
    v_{\rm W,C} = 2.5\times10^{4}\, \,\left( \frac{L_{\rm BH}}{10^{45}\,\rm erg\,s^{-1}} \right)^{0.4}{\rm km\,s^{-1}}.
\label{coldwindvelocity}
\end{equation}
An upper limit of $10^5{\rm km\,s^{-1}}$ is adopted since observations indicate a saturation
of the wind velocity at this value \citep{Gofford:15}.

In the literature, observed value of wind velocity are quite diverse, from the very large value
for ultra-fast outflow to the small value for molecular outflows\citep[e.g.,][]{Crenshaw:03,
Blustin:07, Hamann:08, Tombesi:10}. This is likely because the winds are detected at  different
distances from the black hole: the wind velocity is larger at smaller distance. The diversity of
wind mass flux  is possibly also because of the different detection locations. Specifically, the
mass flux of wind increases with distance because of mass entrainment of the interstellar medium
during the outward propagation of wind. In the observations of \citet{Gofford:15} the winds are
detected at a distance of $\sim 10^{2-4} r_s$ from the black hole. The innermost grid in our
simulation is 2.5 pc (refer to \S\ref{simulationsetup}), which corresponds to $\sim 10^4r_s$
for a typical black hole mass of $2\times 10^9\msun$ in our simulations.  So our adoption of
the observational results of \citet{Gofford:15} is justified.

Eq.~(\ref{eq:coldwindmass}) can be rewritten as
\begin{equation}
        \dot{M}_{\rm W,C} = 0.28 \,\left( \frac{L_{\rm BH}}{L_{\rm Edd}}\right)^{0.85}
                               \left( \frac{L_{\rm Edd}}{10^{45}\,\rm erg\,s^{-1}} \right)^{0.85}\, M_{\odot}\,{\rm yr^{-1}}.
\end{equation}
For the Eddington ratio of $l \equiv L_{\rm BH} / L_{\rm Edd} = 0.1$ and black hole mass
of $M_{\rm BH}=10^{9}\,M_{\odot}$, the  mass flux of wind is $\dot{M}_{W,C}\approx 2.5\,
{M}_{\odot} \,{\rm yr^{-1}}$. This is quite similar to the mass accretion rate of the
black hole, which is $\dot{M}_{\rm BH}=0.1\dot{M}_{\rm Edd}\approx 2\, {M}_{\odot}\,{\rm
yr^{-1}}$.  From eq. (\ref{eq:coldwindenergy}), the power of wind is $\dot{E}_{W,C} = 0.02\,
\dot{M}_{\rm BH}\, c^{2}$. This is five times smaller than the bolometric luminosity, which is
$L_{\rm BH}=0.1\dot{M}_{\rm BH}c^2$.

The above settings of wind properties are different from previous works
\citep[e.g.][]{Novak:11,Gan:14,Ciotti:17}. In these works,  the wind velocity is usually  set to
be a constant value of $v_{\rm W} = 10^{4} \, {\rm km\,s^{-1}}$, while the mass flux or the power
of wind is usually described by a  parameter called  mechanical efficiency ($\epsilon_{\rm W}$),
which is defined as the fraction of mechanical power of wind in the total accretion power. The
value of $\epsilon_W$ is highly uncertain,  its scaling with luminosity is qualitatively
similar to but quantitatively different from what we have just introduced above. Typically,
the wind power described by the above formula is much stronger than that adopted in previous works.

The next question is the spatial (angular) distribution of the mass flux of wind. In reality,
most of mass flux may concentrate close to the surface of the accretion disk. Since our galaxy
model is almost spherically symmetric, and since the relative orientation of the accretion disk
and the galactic disk is random\citep{Schmitt:01},  the exact description of distribution of
wind flux may be not so important so we simply adopt the same description as in previous works
\citep[e.g.,][]{Novak:11, Gan:14, Ciotti:17}. According to this description, the mass flux of
wind $\propto {\rm cos}^2(\theta)$. Thus, the half-opening angle enclosing half of the mechanical
energy is $\approx 45^{\circ}$, and most of the flux concentrates close to the rotation axis
of the accretion disk.

\subsubsection{Radiation}

Since the mass flux of wind is similar to the mass accretion rate of the black hole, the
radiation output from the thin disk can be approximated as,
\begin{equation}
    L_{\rm BH}=\epsilon_{\rm EM,cold} \dot{M}_{\rm BH}c^2
\label{coldflowrad}
\end{equation}
Here $\dot{M}_{\rm BH}$ is the accretion rate close to the black hole calculated by
eq. (\ref{eq:BHaccretionrate}).  For a cold thin disk, the radiative efficiency $\epsilon_{\rm
EM,cold}$ is only a function of black hole spin. In this work we usually set
\begin{equation}
\epsilon_{\rm EM,cold}=0.1,
\label{coldfloweff}
\end{equation}
which means that we assume the black hole is moderately spinning \citep{Wu:13}. To examine
the effect of stronger radiation, sometimes we also consider the case of $\epsilon_{\rm
EM,cold}\approx 0.3$, which corresponds to a rapidly spinning black hole.

The emitted spectrum in the cold accretion mode is represented by the observed multi-waveband
spectrum of quasars. The radiation carries with them energy and momentum. It will heat the ISM
via  photoionization and Compton scattering. It will also be able to push the gas via
electron scattering and photoionization.  The radiative heating and cooling we consider in this
work are computed using the formulae presented in \citet{Sazonov:05}, which describe the
net heating and cooling per unit volume of the gas in photoionization equilibrium. It includes
Compton heating/cooling, bremsstrahlung cooling, photoionization, and line and recombination
cooling. The Compton heating/cooling is usually calculated in terms of ``Compton temperature'',
\begin{equation}
 H_{\rm Compton} = 4.1 \times 10^{-35}n^2\xi  (T_C-T)~{\rm  erg~cm^{-3} s^{-1}},
\label{comptonheating}
\end{equation}
where $\xi=4\pi L_{\rm BH}/n$ is the ionization parameter, $F$ is the flux of photoionizing photons,
$n$ is the number density of the ISM, $T$ is the temperature of the ISM, $T_C$  is the Compton
temperature of the photons, which physically means the energy-weighted average photon energy
of the radiation emitted by the AGN. For the cold feedback mode,  it can be calculated from
the observed spectrum of quasars \citep{Sazonov:04}. The result is
\begin{equation}
T_{\rm C,cold} = 2\times 10^7{\rm K}.
\label{coldtemperature}
\end{equation}


\subsection{Hot accretion (feedback) mode}

Compared with the cold mode, the hot accretion mode corresponds to lower accretion
rates\citep{Yuan:14}, so both the radiation and the wind emitted from this mode are weaker
than those from the cold mode, as shown by Fig. \ref{fig:windradcomp}.  But the hot mode
feedback is still potentially very important. This is because,  compared to the cold mode,
the central AGNs reside in the hot mode for a much longer time than in the cold mode (e.g.,
\citealt{Greene:07}; Figs.~\ref{fig:ldot} \& \ref{fig:dutycycle} of the present work). So we
expect to have an important cumulative effect for the feedback in the hot mode.

\subsubsection{Geometry configuration of accretion flow}

Before we discuss the radiation and wind in the hot accretion mode, we first need to discuss
the geometry of the accretion flow when the accretion rate at the Bondi radius is smaller
than $\dot{M}_{\rm c}$ (Eq. (\ref{criticalrate})). This issue has been well studied in the
cases of the hard state of black hole X-ray binaries and low-luminosity AGNs \citep[see][for a
review]{Yuan:14}. The result is that at large radii, the accretion flow is in the form of thin
disk; but at a certain radius, $r_{\rm tr}$, the thin disk will be truncated and transits into
a hot accretion flow. The value of the transition radius $r_{\rm tr}$ can be described by
(\citealt{Liu:99, Manmoto:00, Gu:00,Yuan:04}; see review by \citealt{Yuan:14})
\begin{equation}
    r_{\rm tr} = 3r_s\left[\frac{2\times 10^{-2}\dot{M}_{\rm Edd}}{\dot{M}(r_{\rm in})}\right]^2.
\label{transitionradius}
\end{equation}
Here $\dot{M}(r_{\rm in})$ is the accretion rate at the innermost grid $r_{\rm in}$ of simulation
calculated by eq. (\ref{mdotbondi}).  The transition radius will be small if $\dot{M}(r_{\rm
in})$ is large. The largest value of $r_{\rm tr}$ is set to be $r_{\rm in}$.

\subsubsection{Wind}
\label{subsubsec:hotwind}

The status of study of wind from hot accretion flows is quite different from the case of
a cold accretion disk.  In this case, the observational data is much fewer compared to the
case of cold disk. This is mainly because the gas in the wind from a  hot accretion flow is
very hot thus generally fully ionized. So it is very difficult to detect them by the usual
absorption-line spectroscopy.  But still, in recent years, we have  gradually accumulated
more and more observational evidences for wind from low-luminosity sources in which we
believe a hot accretion flow is operating, including low-luminosity AGNs and radio galaxies
\citep[e.g.,][]{Crenshaw:12, Tombesi:10, Tombesi:14, Cheung:16}, the supermassive black hole in
our Galactic center, Sgr A* \citep{Wang:13},  and the hard state of black hole X-ray binaries
\citep[e.g.,][]{Homan:16}.  In contrast to the rarity of observational results, we  have
much better theoretical understanding to the wind launched from hot accretion flows, mainly
attributed to the rapid development of numerical simulations in recent years \citep{Yuan:12a,
Narayan:12, Li:13, Yuan:15, Bu:16}. This is also partly because radiation is in general
dynamically not important in hot accretion flow, and technically it is easy to simulate a hot
accretion flow since it is geometrically thick. Especially, the detailed properties of wind from
a hot accretion flow have been carefully studied in \citet{Yuan:15} based on three-dimensional
general relativity MHD simulation data. In this work, to discriminate the turbulent outflow
from real wind, a ``virtual particle trajectory'' approach has been proposed. Compared to the
streamline approach often used in literature, this new approach can give a much more reliable
calculation to the mass flux of wind. In the present work, we will use the results obtained by
\citet{Yuan:15}. We briefly summarize the most relevant results as follows.

We assume that the mass accretion rate at the innermost radius of the simulation domain
$\dot{M}(r_{\rm in})$ is roughly equal to the accretion rate at the outer  boundary of the hot
accretion flow $\dot{M}(r_{\rm tr})$. The outer truncated thin disk should also be able to produce
wind, but in this paper we neglect this part of wind. In this case, the accretion rate close to
the black hole horizon, which determines the emitted luminosity, is  described by \citep{Yuan:15}:
\begin{equation}
\dot{M}_{\rm BH,hot}=\dot{M}(r_{\rm tr})\left(\frac{3r_s}{r_{\rm tr}}\right)^{0.5}\approx \dot{M}(r_{\rm in})\left(\frac{3r_s}{r_{\rm tr}}\right)^{0.5}.
\label{hotaccretionrate}
\end{equation}
So the mass flux of wind launched from the hot accretion flow is described by
\begin{equation}
    \dot{M}_{\rm W,H}\approx \dot{M}(r_{\rm in})\left[1-\left(\frac{3r_s}{r_{\rm tr}}\right)^{0.5}\right].
\label{hotwindflux}
\end{equation}

Compared to jets, the opening angle of wind is much larger (refer to Fig. 1 in \citealt{Yuan:15}),
which makes the interaction between wind and ISM very efficient. According to the detailed
analysis by \citet{Yuan:15} (refer to their Fig. 3),  the mass flux of wind is distributed
within $\theta\sim 30^{\circ}-70^{\circ}$ and $\theta\sim 110^{\circ}-150^{\circ}$   above
and below the equatorial plane respectively. Since the hot accretion flow occupies roughly
$\theta\sim 70^{\circ}-110^{\circ}$, such a distribution implies that the wind is along the
surface of the hot accretion flow. Within the above-mentioned two ranges of $\theta$ angle,
the mass flux of wind is assumed to be independent of $\theta$.


The speed of wind when they just leave the transition radius $r_{\rm tr}$ is $v_{\rm W,H}\approx
0.2 v_{\rm K}(r_{\rm tr})$ (refer to Eq. (7) in \citealt{Yuan:15}).  Here $v_{\rm K}(r_{\rm
tr})$ is the Keplerian velocity at the transition radius $r_{\rm tr}$. When the wind propagates
outward, gravitational force will decelerate it, while gradient force of gas pressure and
magnetic force will accelerate it. The overall result is that  the poloidal velocity of
wind roughly keeps  constant, and its value can be approximated as (\citealt{Yuan:15}, Eq. (8);
Cui, Yuan \& Li 2018, in preparation):
\begin{equation}
v_{\rm W,H}\approx (0.2-0.4) v_{\rm K}(r_{\rm tr}).
\label{windvelocity}
\end{equation}
In this paper we adopt $v_{\rm W,H}=0.2 v_{\rm K}(r_{\rm tr})$. The fluxes of energy and momentum
of wind are described by
\begin{equation}
\dot{E}_{\rm W,H}=\frac{1}{2}\dot{M}_{\rm W,H} v_{\rm W,H}^2,
\label{windpower}
\end{equation}
\begin{equation}
\dot{P}_{\rm W,H}=\frac{2\dot{E}_{\rm W,H}}{v_{\rm W,H}}.
\label{windmomentum}
\end{equation}

The properties of wind described above are quite different from those adopted  in previous
works \citep[e.g.,][]{Novak:11, Gan:14, Ciotti:17}. In these previous works, the velocity of
wind is usually set to be a   value with reference to the observations of wind in the cold
mode. The mass flux or the power of wind  in the hot feedback mode are again determined by the
``mechanical efficiency'' parameter $\epsilon_W$. Its value is now much more uncertain, because
observational constrains to wind in the hot mode are not good. So usually in these works the
wind parameters and their scaling with luminosity are assumed to be the same as those in the
cold mode.  Consequently, we find that the wind power we adopt in the present work  is much
stronger than that in previous works. This causes significant differences of simulation results,
as we will describe later in this paper.

\subsubsection{Radiation}
The radiation output from hot accretion flows has been well studied \citep[see][for a
review]{Yuan:14}. Different from a thin disk, the radiative efficiency not only depends on
the black hole spin, but more importantly on the accretion rate. The radiative efficiency as
a function of accretion rate has been investigated by \citet{Xie:12}. They fit their numerical
calculation results of efficiency by a piecewise power-law function of accretion rate,
\begin{equation}
\epsilon_{\rm EM,hot}(\dot{M}_{\rm BH})=\epsilon_0\left(\frac{\dot{M}_{\rm BH}}{0.1L_{\rm Edd}/c^2}\right)^a,
\label{radefficiency}
\end{equation}
where the values of $\epsilon_0$ and $a$ are given in Table 1 of \citet{Xie:12}.
For convenience, we copy their results here\footnote{The results depend also on a parameter of
the hot accretion flow model $\delta$. In this work we adopt $\delta=0.1$.},
\begin{eqnarray}
   (\epsilon_0, a) &=& \left\{ \begin{array}{ll} (0.2,0.59), & \dot{M}_{\rm BH}/\dot{M}_{\rm Edd}\la 9.4\times 10^{-5} \\
   (0.045,0.27), & 9.4\times 10^{-5} \lesssim \dot{M}_{\rm BH}/\dot{M}_{\rm Edd} \lesssim 5\times 10^{-3} \\
  (0.88,4.53), & 5\times 10^{-3}\lesssim \dot{M}_{\rm BH}/\dot{M}_{\rm Edd} \lesssim 6.6\times 10^{-3} \\
   (0.1,0), & 6.6\times 10^{-3}\lesssim \dot{M}_{\rm BH}/\dot{M}_{\rm Edd} \lesssim 2\times 10^{-2} \end{array} \right.
   \label{efficiencyfit}
\end{eqnarray}
Here $\dot{M}_{\rm Edd}\equiv 10 L_{\rm Edd}/c^2$ is the Eddington accretion rate. Note that the
calculation of \citet{Xie:12} is for a Schwarzschild black hole. Since in the present paper we
assume that the black hole is moderately spinning, we multiply $\epsilon_0$ in \citet{Xie:12}
by a factor of $0.1/0.057$. As shown by \citet{Xie:12},  the radiative efficiency quickly (and
of course also luminosity) increases with the accretion rate, and finally becomes almost equal to
the efficiency of a thin disk at the highest accretion rate of the hot accretion flow, i.e., the
``boundary accretion rate'' between the hot and cold modes, as shown by Fig. \ref{fig:windradcomp}.

The spectrum emitted from a hot accretion flow is quite different from that from a cold thin
disk, e.g., the lack of the big-blue-bump which is present in the typical spectrum of quasars
\citep{Ho:99, Ho:08}. For a given luminosity, the spectrum from a hot accretion flow will
have more hard photons compared to that from a cold disk. This makes the radiative heating
to ISM via Compton scattering in the hot mode more effective than in a cold mode for the same
luminosity. \citet{Xie:17} recently studied this problem in detail and calculated the  Compton
temperature based on the spectral energy distribution of low-luminosity AGNs combined from
literature. The result is
\begin{eqnarray}
    T_{\rm C,hot} &=& \left\{ \begin{array}{ll}
        10^{8}\,{\rm K}, & 10^{-3} \lesssim L_{\rm BH}/L_{\rm Edd} \lesssim 0.02 \\
    5\times 10^{7}\,{\rm K}. ~~& L_{\rm BH}/L_{\rm Edd} \lesssim 10^{-3} \end{array} \right.
    \label{hottemperature}
\end{eqnarray}
This is several times higher than $T_{\rm C,cold}$. This value is smaller than that adopted in
\citet{Gan:14}, in which  they adopt $T_{\rm C,hot}=10^9{\rm K}$ by simple estimation.

\subsection{Comparison of  energy and momentum fluxes of  wind and radiation in cold and hot modes}
\label{subsec:comparison}

\begin{figure*}[!htbp]
    \begin{center}$
        \begin{array}{cc}
            \includegraphics[width=0.45\textwidth]{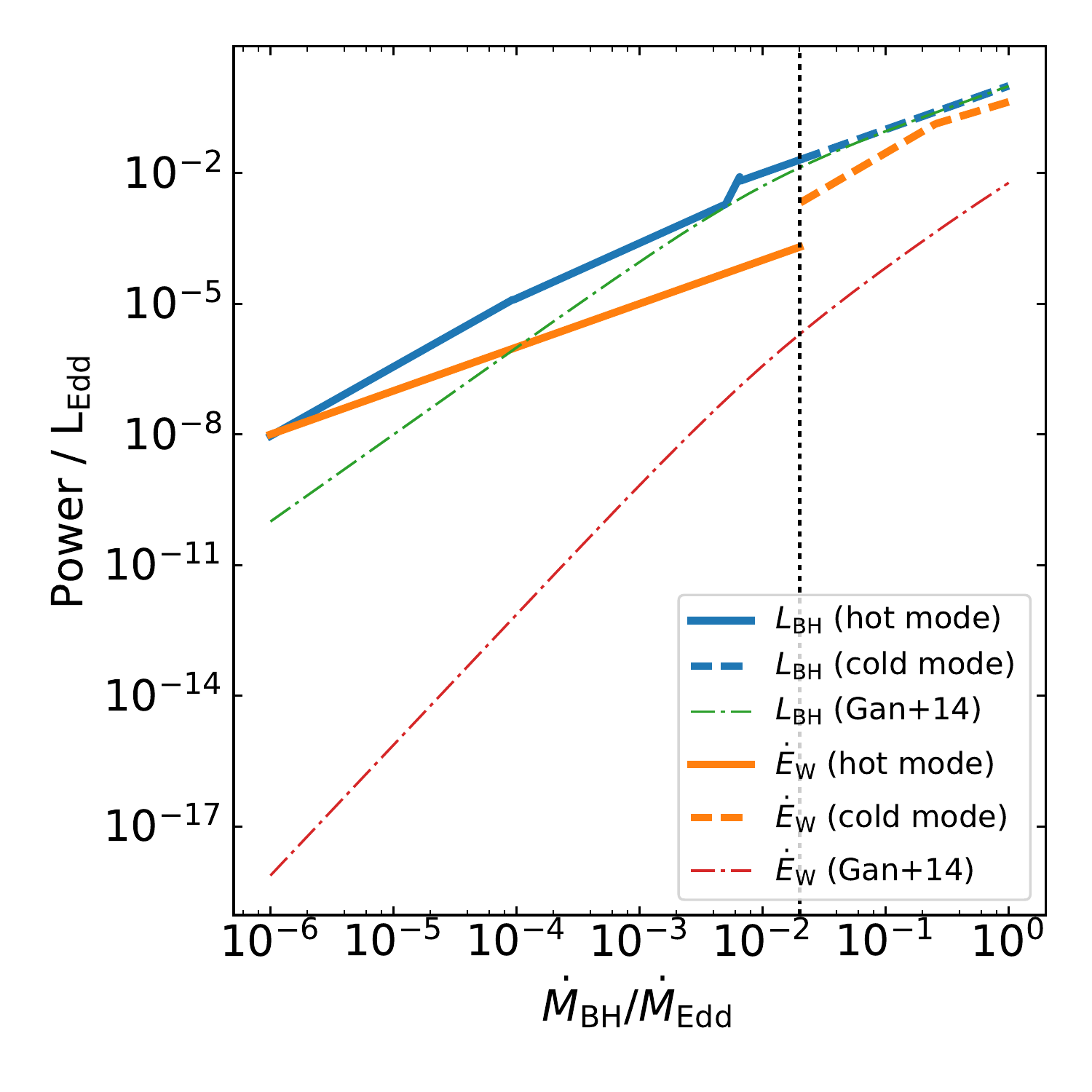} &
            \includegraphics[width=0.45\textwidth]{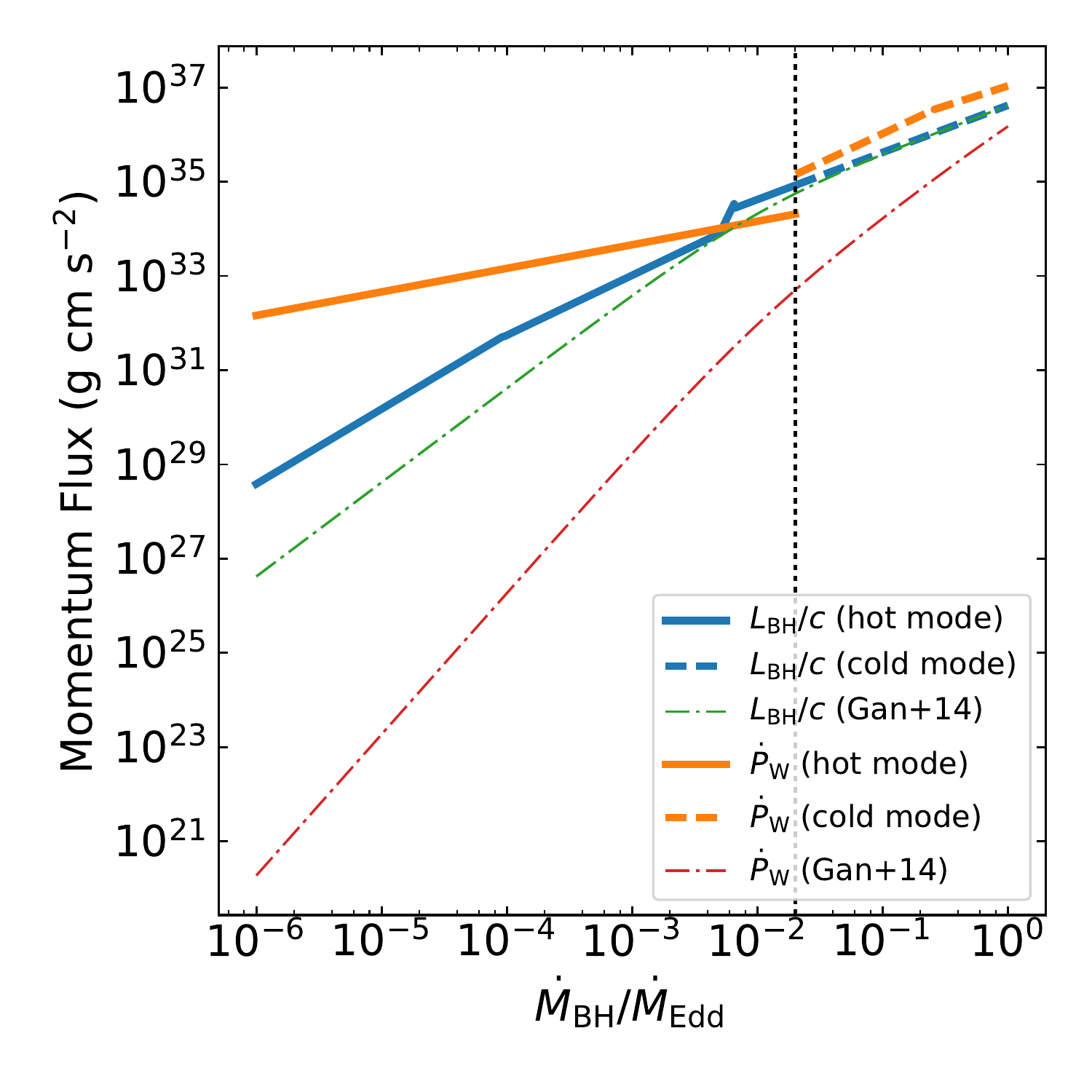}
        \end{array}$
    \end{center}
    \caption{The comparison of the power (left) and momentum flux (right) of wind (orange lines)
    and radiation (blue lines) from hot (solid lines) and cold (dashed lines) accretion modes
    of AGNs.  For comparison, the power and momentum flux of radiation and wind from \citet{Gan:14}
    are also shown by the dot-dashed lines. } 
    \label{fig:windradcomp}
\end{figure*}

We now compare the fluxes of energy and momentum of wind and radiation in both cold
and hot feedback modes. For the cold feedback mode, the energy and momentum fluxes of
radiation can be obtained by Eq. (\ref{coldflowrad}) and $L_{\rm BH}/c$, respectively. The
energy and momentum fluxes of wind can be calculated from eqs. (\ref{eq:coldwindenergy}) and
(\ref{eq:coldwindmom}). For the hot feedback mode, the energy and momentum fluxes of wind  can be
calculated from eqs. (\ref{windpower}) \& (\ref{windmomentum}). The wind velocity $v_{\rm W,H}$
can be obtained from eq. (\ref{windvelocity}) while the value of transition radius $r_{\rm tr}$
can be obtained by combining eqs. (\ref{transitionradius}) \& (\ref{mdotbondi}). The  energy
and momentum fluxes of radiation are $L_{\rm BH}=\epsilon_{\rm EM,hot}\dot{M}_{\rm BH}c^2$
and $L_{\rm BH}/c$, respectively. The value of $\epsilon_{\rm EM,hot}$ can be calculated from
eqs. (\ref{radefficiency}) \& (\ref{efficiencyfit}).

The results of such a comparison are shown in Fig. \ref{fig:windradcomp}. The left and right
panels denote the power and momentum fluxes, respectively.  In the hot mode, it is somewhat
surprising to note that the power of radiation is larger than that of wind.  This is because
radiation comes from the innermost region of the accretion flow where gravitational energy release
is the largest, while wind production is suppressed at that region due to the suppression of
turbulence \citep{Yuan:15}.  But on the other hand,  the momentum flux of wind is in general
larger than that of radiation. This is consistent with the fact that wind production in the
hot accretion flow is not driven by radiation, but by the combination of magnetic  and thermal
mechanisms \citep{Yuan:15}.  It is interesting to note that in the cold accretion mode, the
momentum flux of wind is also significantly larger than that of radiation. Generally we think
the wind in the cold mode is mainly driven by radiation. This result indicates that magnetic
field likely also plays an important role. The large momentum flux of wind in both the hot
and cold modes suggests that wind will be important in pushing the gas surrounding the AGN
away. This is confirmed by detailed simulation as we will describe later in this work.

In order to see clearly the differences between our current descriptions of radiation and
wind and those in previous works, we have also shown the AGN model from \citet{Gan:14}
in Fig.~\ref{fig:windradcomp}.  We can see from the figure that in the hot mode, both the
radiation and wind are much stronger in the present work than in \citet{Gan:14} in terms of
both power and momentum flux. In the cold mode, the wind adopted in the current work is also
significantly stronger than in \citet{Gan:14}, but the radiation almost remains unchanged.

We note that because of the intrinsic difference between radiation and wind, especially when
they interact with the ISM, we should not judge which one is more important in the feedback
simply from their magnitude of power and momentum flux. This is because the cross section
of photon-particle and particle-particle interaction differs by orders of magnitude, so it will
take very different distances for wind and radiation to convert their energy and momentum to the
ISM. Let's now estimate their ``typical length-scale of feedback'', $l_{\rm rad}$ and $l_{\rm
wind}$, which we define as the distance within which they can transport a significant fraction
of their momentum or power to the ISM. This is basically the mean free path of photons and
wind particles.  Note that these are only low limits of the spatial range within which  radiation
or wind can affect the ISM. The real range should be much larger, especially for wind. For
radiation, $l_{\rm rad}$  is the distance where the scattering optical depth is equal to one,
i.e., $\sigma_{T}\rho l_{\rm rad}/m_p=1$,  here $m_p$ is the proton mass and $\sigma_T=6.65\times
10^{-25} {\rm cm}^2$ is the Thompson electron scattering cross section. So we have
\begin{equation}
l_{\rm rad}\sim \frac{m_p}{\sigma_T\rho}\sim 10  {\rho}_{-24}^{-1}~{\rm kpc}.
\label{radlength}
\end{equation}
Here $\rho_{-24}\equiv \rho/10^{-24}{\rm g}{\rm cm}^{-3}$, the typical mass density in the central
region of the galaxy is assumed to be $10^{-24}{\rm g}{\rm cm}^{-3}$ (refer to the right panel
of Fig. \ref{fig:agnoutburst}). We only consider Thompson electron scattering when estimating
the value of $l_{\rm rad}$. Its value will be smaller when line absorption is taken into account.

For interaction between wind and ISM, the cross section due to Coulomb collision is $\sigma_C\sim
\pi r_e^2\sim \pi e^4/k^2T^2\sim 10^{-4}T^{-2}{\rm cm}^{2}$, with $e$ being the electron charge,
$k$ the Boltzmann constant, and $T$ typical temperature of ISM \citep[e.g.,][]{Ogilvie:16}. So we have
\begin{equation}
 l_{\rm wind}\sim \frac{m_p}{\sigma_C \rho}\sim 0.5{\rho}_{-24}^{-1}T_{7}^2~{\rm pc}.
 \label{windlength}
\end{equation}
Here $T_7\equiv T/10^7K$.  The typical length scale of feedback for wind is much smaller
than that for radiation for typical parameters of our problem. This is because the Coulomb
cross section is orders of magnitude larger than the Thompson scattering cross section,
$\sigma_C\gg\sigma_T$. This result indicates that wind can  more easily  deposit its energy and
momentum to the ISM than radiation. So we expect that wind should be in general more effective
to control the mass accretion rate of the black hole than radiation since the accretion rate
is determined mainly by the properties of the gas very close to the black hole. This suggests
we should especially pay more attention to the role of wind in the hot feedback mode, because
as we will see from Fig. \ref{fig:ldot}, the AGN spends most of its time in the hot  mode.
In fact, the important role of wind from hot feedback mode has begun to be gradually recognized
by researchers. One such an example is a most recent work by \citet{Weinberger:17a}. In this
important work, they invoke wind from the hot accretion flow to achieve a sufficiently rapid
reddening of moderately massive galaxies without expelling too many baryons.

But radiative feedback potentially has its ``advantage'' compared to the feedback by wind. Most
importantly, radiation is more powerful than wind by a factor of few in the cold mode, as we can
see from Fig. \ref{fig:windradcomp}. However, how efficiently the radiation can deposit its energy
to the ISM in the galaxy, or what fraction of the radiation energy can be converted to the ISM,
depends also on the optical depth of the galaxy. In our current model, from the right panel of
Fig. \ref{fig:agnoutburst}, the spatially-averaged mass density is $\sim (10^{-25}-10^{-26}){\rm
g}{\rm cm}^{-3}$, so the scattering optical depth of the whole galaxy is
\begin{equation}
\tau\sim \sigma_T/m_p l_{\rm galaxy}\sim (0.01-0.1).
 \end{equation}
This implies that  the radiation from the AGN can only deposit $\sim (1-10)\%$ of its energy to
the ISM of the host galaxy. This suggests that radiation feedback may be less important compared
to wind in our present work. As we will see, this is confirmed by simulations in the present
work. We would like to point out two caveats here.  One is that in the present paper we only
consider an isolated galaxy. If we take into account external gas supply, or more generally if
we consider some more gas-rich galaxy, the density of the ISM will increase so the optical depth
of the galaxy will become much larger, therefore radiative feedback will become more important.
Another caveat is that in the current work, we ignore the effect of dust in the ISM. If the
dust were included, the opacity could be orders of magnitude larger than the electron-scattering
opacity \citep{Novak:12}, thus a much larger portion of radiation could be deposited to the ISM.

One characteristic output of LLAGNs in the hot accretion mode is jets \citep{Yuan:14}. The
comparison between the jet power and the radiative power has been done in the case of the
hard state of black hole X-ray binaries based on observational data \citep{Fender:03}.  It was
found that  the radiation power  is larger than the jet power when  the luminosity is not too
low, $L_{\rm BH}\ga (10^{-3}\sim 10^{-4})L_{\rm Edd}$. In the theoretical aspect, \citet{Yuan:15} have
compared the power and momentum flux between jet and wind. However, since that work deals with
a non-spinning black hole, only the disk-jet, which is powered by the rotation of the accretion
flow rather than the black hole spin, is considered; while the Blandford-Znajek jet is not. In
that case, it is found that both the power and the momentum flux of wind is much larger than
that of jet. In the case of a rapidly spinning black hole, it is expected that the  jet power
will be much larger. It is unclear how the comparison between jet and wind (and radiation) will
change and this requires future investigation (Yuan et al. in preparation). In addition to the
direct comparison of their power, another important factor when we compare their importance of
feedback is their coupling efficiency with the ISM. As we argue above, it looks that only a small
fraction of the radiation power is converted to the ISM. It is desirable to study the case of jet.

\section{Model}\label{sec:Model}

In this section, we discuss the other aspects of the model,  such as the galaxy
model, the treatment of stellar evolution, and the hydrodynamical equations. These  are same as in
\citet{Gan:14}. For completeness, here we briefly describe them as follows.

\subsection{Model of galaxy and stellar evolution }

Our galaxy model refers to an isolated elliptical galaxy. The gravitational potential consists
of the contributions by a dark matter halo, a stellar spheroid embedded in it, plus a central
black hole.  They dominate the gravity beyond 10 ${\rm kpc}$, $0.1-10~{\rm kpc}$, and within
$0.1 ~{\rm kpc}$, respectively. The self-gravity of ISM is ignored in our simulation.

For consistency with previous works and ease of comparison, and for simplicity, we assume that
the galaxy evolves in isolation without any external fuel source, either from accretion from the
intergalactic medium or from acquisition by mergers. Following previous works, we also neglect the
initial ISM in our work, and all the gas resource fueling the black hole comes from the stellar
evolution, including stellar wind and supernovae. This is an important caveat of our present work.
In this case, some results shown are simply for qualitative illustration, not rigorous comparison
with observations. In the future we will include the effects of external gas supply and mergers.

The calculation of the stellar evolution in our simulation follows the description presented in
\citet{Ciotti:12}. In fact, over a cosmological time span, the total stellar wind injected can
reach $20\% \sim 30\%$ of the total initial stellar mass, which is two orders of magnitude larger
than the black hole mass. Both the stellar wind and supernova explosion will provide  sources
of mass and energy into the galaxy and these effects will be taken into account in
our simulations. This gas, when it cools due to radiation, will form stars.  Some newly formed
massive stars evolve quickly and explode via Type II supernovae. These processes are considered
in the simulation and their calculations are described in \S~\ref{subsec:starformation}.

The stellar distribution is described by the Jaffe profile \citep{Jaffe:83},
\begin{equation}
    \rho_{\star} = \frac{M_{\star}\,r_{\star}}{4\pi r^{2} (r_{\star}+r)^{2}}
    \label{stellardis}
\end{equation}
where $M_{\star}$ is the total stellar mass, and $r_{\star}$ is the scale length of the
galaxy, which corresponds to the projected half-mass radius (i.e., effective radius) of
$R_{e} = 0.7447\,r_{\star} = 6.9$ kpc \citep{Ciotti:09}.
The density profile of the dark matter halo is set so that the total mass profile
decreases as $r^{-2}$, as observed \citep[e.g.,][]{Rusin:05, Czoske:08, Dye:08}. The values of
model parameters are chosen so that the galaxy obeys the edge-on view of the fundamental plane
\citep{Djorgovski:87} and the Faber-Jackson relation \citep{Faber:76}. The total stellar mass
$M_\star=3\times 10^{11}\msun$,
the velocity dispersion is set to be $\sigma = 260\,{\rm km\,s^{-1}}$, the stellar mass-to-light
ratio is $M_{\star}/L_{\rm B}=5.8$, where the total $B$-band luminosity is $L_{\rm B} =
5\times 10^{10} \,L_{\rm B\odot}$.  The initial black hole mass is determined according to
the correlation between the black hole mass and galaxy mass \citep[e.g.,][]{Magorrian:98,
Kormendy:13}. In this paper, we adopt the more updated correlation given in \citet{Kormendy:13},
which gives the initial mass of the black hole of $M_{\rm BH}=6\times 10^{-3}M_\star$ for
$M_{\star}=3\times 10^{11}\msun$. But simply to examine the effect of the black hole mass,
sometimes we also run several models using the ``old'' \citep{Magorrian:98} correlation which
gives $M_{\rm BH}=10^{-3}M_\star$ for comparison purpose.

Most of gas is provided by stellar evolution in our work. So the initial angular momentum of
gas ejected from the star is determined by the stellar rotation. In this paper, we assume that
the stars rotate slowly.  The rotation profile is described by \citet{Novak:11},
\begin{equation}
    \frac{1}{v_{\phi}(R)} = \frac{d}{\sigma_{0}\,R} + \frac{1}{f\sigma_{0}} + \frac{R}{j},
\end{equation}
where $v_{\phi}$ is the rotation velocity, $R$ is the distance to the $z$-axis, and $\sigma_{0}$
is the central one-dimensional line-of-sight velocity dispersion for the galaxy model. Here,
$d,f,$ and $j$ are parameters that control the angular momentum profile. In this model,
the stars rotate with solid body at $R<d$ and with constant specific angular momentum of $j$
at larger radii. We adjust the parameter to avoid forming a rotationally supported gas disk
inside the innermost grid cell of our simulation domain. In the companion paper \citep{Yoon:18},
we will consider the high angular momentum case.

\subsection{Energy and momentum interaction between radiation and ISM}
\label{subsec:radiativefeedback}

The radiation emitted from the central AGN will heat or cool the ISM, and also exerts a radiation
force to the ISM.  We calculate the radiative heating and cooling based on the formulae
presented in \citet{Sazonov:05}, which describe the net heating or cooling rate per unit volume
of a cosmic plasma in photoionization equilibrium. The processes considered include Compton
heating and cooling, bremsstrahlung loss, photoionization, line and recombination heating
and cooling. In particular, the calculation of the Compton heating or cooling is described
by eq. (\ref{comptonheating}) in terms of the Compton temperature. As we emphasize in
\S~\ref{sec:agnphysics}, since the typical SED  emitted in the cold and hot accretion
(feedback) modes are very different, the corresponding Compton temperature in the hot mode
is several times higher than that in the cold mode (refer to eqs. (\ref{coldtemperature}) and
(\ref{hottemperature})). Some simplifications are adopted. We neglect the effect of dust. The
radiative transport is also considered, but in an approximated way by assuming the flow is
optically thin.

For the momentum interaction, i.e., the radiation force, we follow \citet{Novak:11} and
consider both the electron scattering and the absorption of photons by atomic lines. The former
is described by
\begin{equation}
(\nabla p_{\rm rad})_{\rm es}=-\frac{\rho \kappa_{\rm es}}{c}\frac{L_{\rm BH}}{4\pi r^2}.
\end{equation}
Here $\kappa_{\rm es}=0.35~{\rm cm}^2~{\rm g}^{-1}$ is the electron scattering opacity. The
latter is described by
\begin{equation}
(\nabla p_{\rm rad})_{\rm photo}=\frac{H}{c}.
\end{equation}
Here $H$ is the radiative heating rate per unit volume.

\subsection{Star formation}
\label{subsec:starformation}

Star formation is implemented by subtracting mass, momentum and energy from the grid (see
\citealt{Novak:11} for details). The  star formation rate per unit volume is determined by
\begin{equation}
    \dot{\rho}_{\rm SF} = \frac{\eta_{\rm SF}\,\rho}{\tau_{\rm SF}},
\end{equation}
where we adopt the SF efficiency of $\eta_{\rm SF}=0.1$, and the SF time scale,
$\tau_{\rm SF}$ is
\begin{equation}
    \tau_{\rm SF} = \max(\tau_{\rm cool},\tau_{\rm dyn}),
\end{equation}
where the cooling time scale, $\tau_{\rm cool}$, and the dynamical time scale, $\tau_{\rm dyn}$,
are
\begin{equation}
    \tau_{\rm cool} = \frac{E}{C},~~ \tau_{\rm dyn}=\min(\tau_{\rm ff},\tau_{\rm rot})
\end{equation}
with
\begin{equation}
    \tau_{\rm ff} = \sqrt{\frac{3\,\pi}{32G\rho}},~~ \tau_{\rm rot}=\sqrt{r\frac{\partial\,\Phi(r)}{\partial\,r}},
\end{equation}
where $E$ is the internal energy density, $C$ is the cooling rate per unit volume, and $\Phi(r)$
is the gravitational potential at a given radius.

The corresponding loss rates of energy and momentum due to star formation are
\begin{equation}
\dot{E}_{\rm SF}=\frac{\eta_{\rm SF}E}{\tau_{\rm SF}}, ~~~~\dot{\mathbf{m}}_{\rm SF}=\frac{\eta_{\rm SF}\mathbf{m}}{\tau_{\rm SF}}=\dot{\rho}_{\rm SF} \mathbf{v}
\end{equation}
Here $\mathbf{m}$ is the momentum density of the ISM and $\mathbf{v}$ is the velocity vector
of the ISM.

On the other hand, among the newly formed stars, there is a population of massive stars. The
massive stars have a relatively short lifetime and will finally evolve to Type II supernovae
(SN II) on a relatively short  timescale. They will then eject mass and energy into ISM at some
rates. This has also been considered in our simulation.  We note that there is a caveat in our
simulations that we do not take into account the migration of stars; instead they keep their
location all the time.

\subsection{Hydrodynamics}

The evolution of the galactic gas flow, given all the above physical processes including star
formation and AGN feedback,  is described by the following time-dependent Eulerian equations
for mass, momentum, and energy conservations \citep[e.g.,][]{Ciotti:12}:
\begin{equation}
   \frac{\partial \rho}{\partial t} + \nabla \cdot \left( \rho \mathbf{v} \right) = \alpha\,\rho_{\star} + \dot{\rho}_{II}-\dot{\rho}_{\star}^{+},
\end{equation}
\begin{equation}
    \frac{\partial \mathbf{m}}{\partial t} + \nabla \cdot \left( \mathbf{m v} \right) = -\nabla p_{\rm gas} + \rho\mathbf{g} -\nabla p_{\rm rad} - \dot{\mathbf{m}}_{\star}^{+},
\end{equation}
\begin{equation}
    \frac{\partial E}{\partial t} + \nabla \cdot \left( E \mathbf{v} \right) = -p_{\rm gas}\nabla \cdot \mathbf{v} + H - C + \dot{E}_{S} + \dot{E}_{I} + \dot{E}_{II} - \dot{E}_{\star}^{+},
\end{equation}
where $\rho,\,\mathbf{m},$ and $E$ are the gas mass, momentum and internal energy per unit volume,
respectively. $\mathbf {v}$ is the velocity, $p_{\rm gas}= (\gamma -1)E$ is the gas pressure,
the adiabatic index $\gamma = 5/3$, and $\mathbf {g}$ is the gravitational field of the galaxy
(i.e., stars, dark matter, plus the time-dependent contribution of the growing central SMBH).
$\alpha\,\rho_{\star}$ is the mass source from the stellar evolution,  $\dot{E}_{\rm S}$
corresponds to the thermalization of the stellar wind due to stellar velocity dispersion,
 as the ejected gas collides with the mass lost from other stars and/or with the ambient  gas
\citep{Parriott:08}. This process provides heat to the ISM at a rate, $\dot{E}_{\rm S} =
1/2\,\alpha\,\rho_{\star} \rm Tr(\sigma^{2})$, where  $\sigma$ is the isotropic one-dimensional
stellar velocity dispersion without the contribution of the central black hole \citep{Ciotti:09}.
\begin{equation}
    \sigma^{2}(s) = \sigma_{0}^{2}(1+s)^{2}s^{2}
                    \left[ 6 \ln\left( \frac{1+s}{s} \right) + \frac{1-3s-6s^{2}}{s^{2}(1+s)} \right],
\end{equation}
where $s\equiv r/r_{\star}$.
The term $\dot{\rho}_{II}$ in the mass equation denotes the mass return from SNe II, while
$\dot{\rho}^+_*$, $\mathbf{m}^+_*$ and $\dot{E}^+_*$ denote the sink terms of mass, momentum, and
energy due to star formation, respectively. In the energy equation, $\dot{E}_I$ and $\dot{E}_{II}$
are the feedback rates of energy from SNe I and SNe II, respectively. Finally, $H$ and $C$
denote the radiative heating and cooling rates (\S~\ref{subsec:radiativefeedback}). So totally
we have three different heating mechanisms, i.e,. AGN heating, stellar heating, and supernova
heating. It is then an interesting question which one dominates over the others. This is the
topic of another paper \citep{Li:18}. We find that the answer depends on the
region in the galaxy, the time, and the properties of galaxy.  Roughly speaking, stellar heating
processes ($\dot{E}_{S},\, \dot{E}_{I},\,{\rm and}\, \dot{E}_{II}$) likely dominates over the AGN
heating at the galactic outskirt, while supernova heating is more important than stellar heating.

\subsection{Simulation Setup}
\label{simulationsetup}

We employ the parallel ZEUS-MP/2 code \citep{Hayes:06}, using two dimensional axisymmetric
spherical coordinates ($r,\theta, \phi$). Following \citet{Novak:11}, in the $\theta$ direction
the mesh is divided homogeneously into 30 grids; while in the radial direction, covering the
radial range of 2.5 pc -- 250 kpc,  we use a logarithmic mesh with 120 grids.  A small range of
$\theta$ around the axis is excluded to avoid singularity there.  With such  grids, the finest
resolution is at the inner-most grid, which is $\sim$ 0.3 pc. Such kind of configuration is
obviously essential since the innermost region is the place where radiation and wind from AGN
originate, and thus most important.  In particular, the inner boundary radius is chosen to resolve
the Bondi radius.  For the gas with sound speed of $c_s$, the Bondi radius is estimated to be
\begin{equation}
r_B=\frac{GM_{\rm BH}}{c_s^2}.
\end{equation}
In the general case, the accretion flow is not homogeneous but mixture of cold clumps and
hot gas. The highest temperature the gas can reach for the hot phase gas can be roughly
estimated by the Compton temperature $T_{\rm C}$.  In the hot feedback mode, we have $T_{\rm
C,hot}\approx 10^8{\rm K}$. Considering a typical black hole mass of $M_{\rm BH}=2\times 10^9\msun$
(refer to Fig. \ref{fig:bhmass}), we have $r_B=6 {\rm pc}$, which is larger than the radius
of the inner boundary of our simulation domain (2.5 pc). For the cold mode, the Bondi radius will be
even larger thus more easily resolved.

The accretion rate at the innermost radius of the simulation $r_{\rm in}$ is calculated by,
\begin{equation}
     \dot{M}(r_{\rm in}) = 2\pi r_{\rm in}^2\int^{\pi}_0 \rho(r_{\rm in},\theta) ~ {\rm min} \left[v_{r}(r_{\rm in},\theta),0\right] \sin{\theta}\,d\theta.
\label{mdotbondi}
\end{equation}
Note that both the hot and cold phases of the gas are included in the above calculation.
Such a calculation of accretion rate is obviously much more precise than that given by  Bondi
accretion rate formula often adopted in literature, which assumes an accretion of single-phase
and non-rotating gas (see \citealt{Negri:17} for a summary of the problems of using the simple
Bondi accretion rate formula; see also \citealt{Gaspari:18} for the discussion of ``chaotic
cold accretion''.).

As for the boundary condition, in the inner and outer radial boundary we use the standard
``outflow boundary condition'' in the ZEUS code (see \citealt{Stone:92} for more details), so
that the gas is free to flow in and out at the boundary. For $\theta$ direction, a ``reflecting
boundary condition'' is set at each pole. A  temperature floor of $10^4{\rm K}$ is adopted in
the cooling functions, since the gas cannot reach these low temperatures by radiative cooling
alone \citep{Sazonov:05, Novak:11}.

\begin{deluxetable*}{cccccccc}[!htbp]
    \tablecolumns{8}\tabletypesize{\scriptsize}\tablewidth{0pt}
    \tablecaption{Description of the simulations \label{tab:model}}
    \tablehead{ \colhead{model} & $M_{\rm BH}(\msun)$ & $\epsilon_{\rm EM,max}$ & \colhead{Mechanical} & \colhead{Radiative}
              & \colhead{$M_{\rm BH,final} (\msun)$} & \colhead{$\Delta M_{\rm \star,final} (\msun)$} & \colhead{duty cycle (\%)} \\
                \colhead{}      & \colhead{}          & \colhead{}              & \colhead{Feedback} & \colhead{Feedback} & \colhead{} & \colhead{} & \colhead{}}
    \startdata
        fullFB     & $1.8\times 10^9$ & 0.1 & o & o & $2.1\times 10^9$    & $6.5\times 10^9$    & $2.4\times10^{-2}$ \\
        windFB     & $1.8\times 10^9$ & 0.1 & o & x & $2\times 10^9$      & $6.6\times 10^9$    & $2.6\times10^{-2}$ \\
        radFB      & $1.8\times 10^9$ & 0.1 & x & o & $1.8\times 10^{10}$ & $1\times 10^{10}$   & 7.4 \\
        noFB       & $1.8\times 10^9$ & 0.1 & x & x & $1.3\times 10^{10}$ & $1.2\times 10^{10}$ & -- \\
        fullFBem03 & $1.8\times 10^9$ & 0.3 & o & o &  $1.9\times 10^{9}$ & $5.8\times 10^{9}$  & $9.9\times10^{-3}$ \\
        windFBem03 & $1.8\times 10^9$ & 0.3 & o & x &  $1.8\times 10^{9}$ & $7.2\times 10^{9}$  & $9.8\times10^{-3}$ \\
        fullFBmag  & $3\times 10^8$   & 0.1 & o & o & $3.3\times 10^{8}$  & $8.6\times 10^{9}$  & $3.8\times10^{-2}$ \\
    \enddata
\end{deluxetable*}

~~~~~
\section{Results}
\label{sec:Results}

\begin{figure*}[!htbp]
    \begin{center}$
        \begin{array}{cc}
            \includegraphics[width=0.45\textwidth]{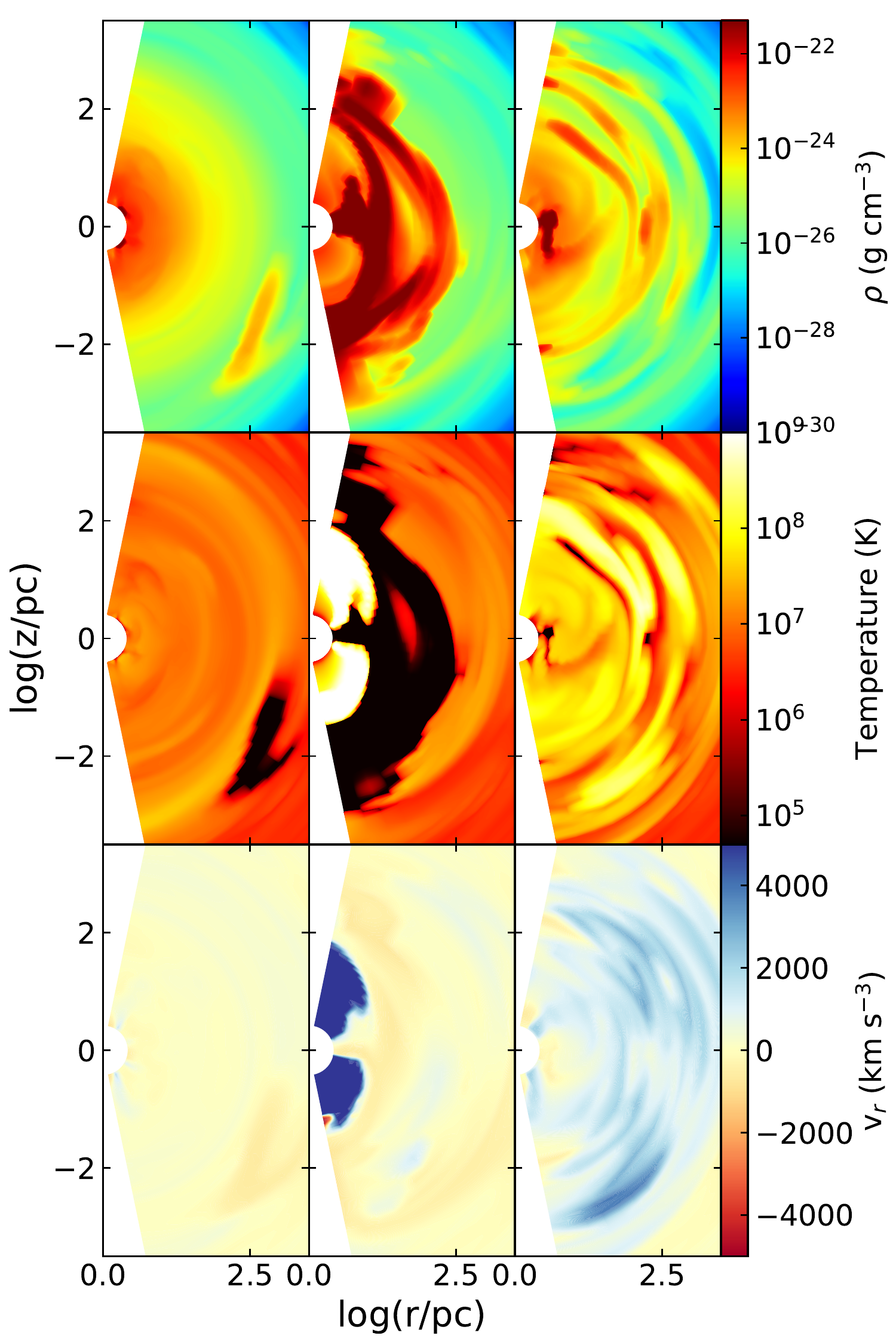} &
            \includegraphics[width=0.45\textwidth]{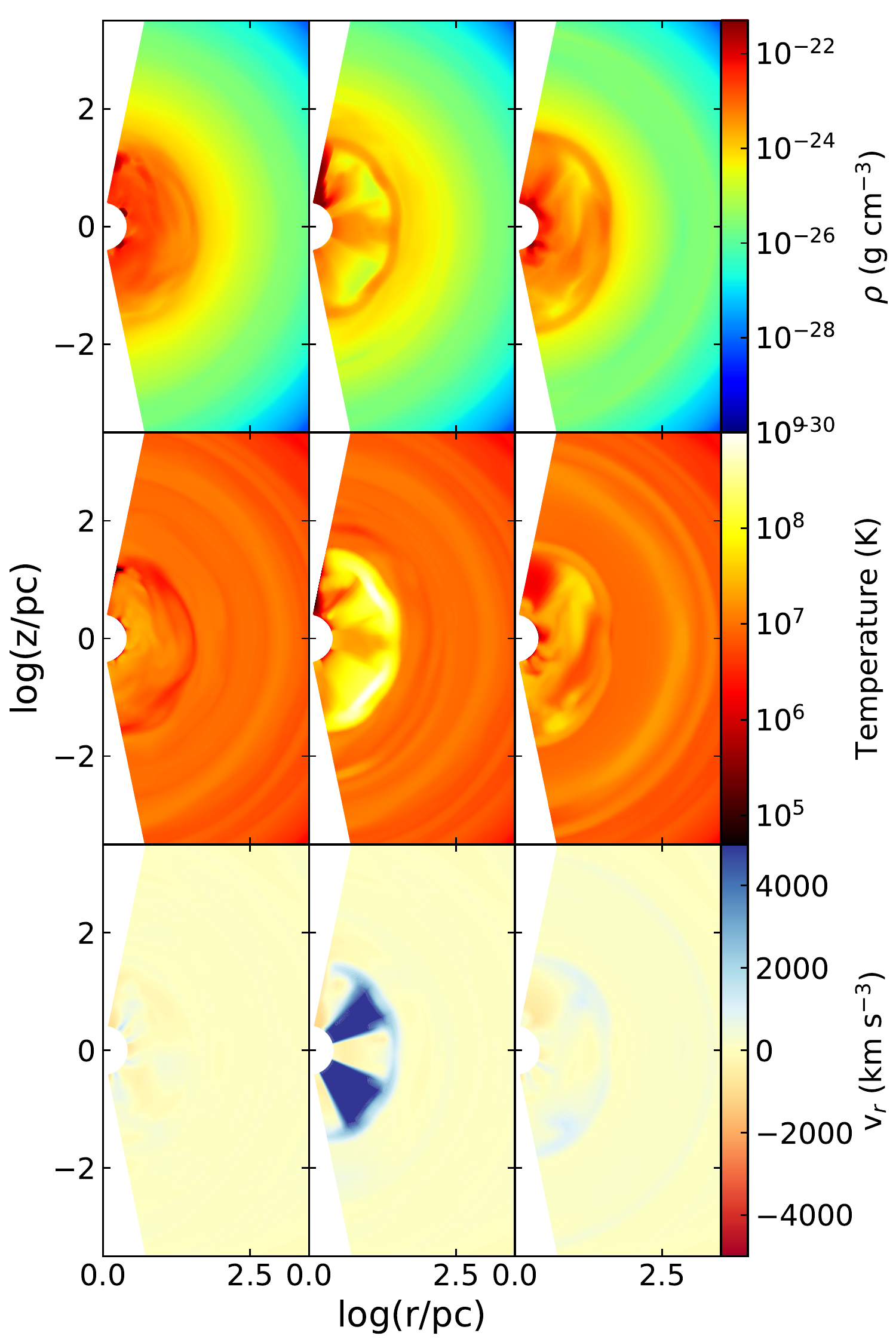}
        \end{array}$
    \end{center}
    \caption{Spatial distribution of density (upper), temperature (middle), and radial velocity (bottom) at three different
    times in correspondence with  an outburst of the AGN. The left column is for an
    outburst occurring in the cold feedback mode. The left, middle, and right plots correspond
    to $t=1.516$ Gyr (immediately before the outburst; density and temperature are both quite
    smooth in the central region of the galaxy), 1.525 Gyr (close to the peak of the outburst;
    two outflowing low-density wind region are evident in the two polar directions, up to
     100 pc; dense and low-temperature gas around the equatorial plane is inflowing to
    fuel the black hole), and 1.54 Gyr (just after the outburst; the density and temperature
    become smooth again in the central region of the galaxy, but different from the epoch
    before outburst, the gas in most of the central region of the galaxy is still outflowing.),
    respectively. The right column is for an outburst occurring in the hot feedback mode. The
    left, middle, and right plots correspond to $t=1.815$ Gyr (immediately before the outburst;
    in the central region of the galaxy density and temperature are less smooth compared to
    the cold-mode outburst in the left column, and more gas is outflowing), 1.82 Gyr (close
    to the peak of the outburst; compared to the cold-mode outburst in the left column, in
     two outflowing wind region are also evident but weaker;  while
    around the equatorial plane,  the gas is also  inflowing to fuel the black hole, but the
    gas is less dense and temperature is higher), and
    1.83 Gyr (just after the outburst; compared to the cold-mode outburst in the left column,
    the density and temperature also become smooth again in the central region of the galaxy,
    but  the gas is mainly inflowing), respectively.\\  }
    \label{fig:agnoutburst}
\end{figure*}

In this work, we consider both the cold and hot feedback modes, and in each mode the feedback
by radiation and wind are  taken into account. In order to understand the respective roles of
radiation and wind,  we carry out four runs: one with both mechanical and radiative feedbacks
(fullFB), one with only mechanical feedback (windFB), one with only radiative feedback
(radFB), and the last one with no feedback (noFB).  In addition, we also perform a run with
higher radiative efficiency of $\epsilon_{\rm EM}=0.3$ (fullFBem03), which corresponds to
the case of a rapidly spinning black hole. The model with a smaller initial black hole mass
based on the \citet{Magorrian:98} correlation is also calculated for comparison and denoted as
``fullFBmag''. All these models are listed in Table~\ref{tab:model}.  The final values of black
hole mass, the accumulated mass of new stars, and duty cycle (the ratio of the  duration in
the cold mode and the total duration of AGN) have also been given in the table.

In order to investigate the effects of different  AGN physics, we will compare our results
with relevant previous works by \citet{Gan:14} and \citet{Ciotti:17}. The model framework of
these two works are very similar to our present paper, except the AGN physics in both the cold
and hot feedback modes.  We specifically choose to compare our results with the model ``B05v''
in \citet{Gan:14}.

\subsection{Overview of the Evolution}
\label{overallscenario}

Our simulation starts from an age of 2 Gyr of the stellar population. If there were no AGN
feedback, the galaxy would evolve smoothly.  When AGN feedback is included, the overall evolution
of the galaxy is similar to previous works \citep[e.g.,][]{Novak:11, Gan:14, Ciotti:17}. In
this case,  on the one hand, the radiation and wind from the AGN will interact with the gas
in the galaxy and change their properties, especially the spatial distributions of density
and temperature as we will see from Fig. \ref{fig:agnoutburst}.  The changes of density and
temperature will subsequently result in the change of star formation and the whole evolution
of the galaxy. On the other hand, the change of the properties of the gas will also affect the
fueling and  activity  of the AGN and the black hole growth. Especially, the activity of the
AGN will strongly fluctuate, as we will explain in the following paragraphs. This results in the
duty-cycle of AGN. In the following subsections, we will discuss these issues one by one.

In this subsection, we focus on introducing the general scenario of the evolution of AGN
activity and the feedback effects on the gas in the galaxy.  For this aim, we have drawn
Fig. \ref{fig:agnoutburst}. There are two columns in this figure, with the left and right
one corresponding to an outburst occurred  in the cold mode (left) and the hot mode (right),
respectively. In each column, from top to bottom, we show the evolution of density, temperature,
and radial velocity of the gas in the galaxy before, during, and after the outburst.  The three
plots in the left column correspond to $t=1.516$ Gyr (immediately before the outburst), 1.525 Gyr
(close to the peak of the outburst), and 1.54 Gyr (just after the outburst), respectively. The
maximum  accretion rate in this interval can reach 0.41$\dot{M}_{\rm Edd}$.  Before the outburst,
since the accretion rate of AGN is low, the radiation and wind from the central AGN are weak,
thus the galaxy is hardly disturbed. So we can see from the figure that, in the central region
of the galaxy, $r\la 100$ pc, the spatial distributions of both density and temperature of the
gas are quite smooth, and the gas are all inflowing toward the black hole. These gas will cool
by radiation so density will become higher. We can see from the figure that there are many cold
dense clumps and filaments outside of $\sim 100$ pc. They are formed by thermal instability
and Rayleigh-Taylor instability of the gas. They are obviously the ideal place of star formation.
The fall of these clumps  will significantly increase the accretion rate of the black hole  and
causes the AGN to enter into the outburst phase on a timescale of $\sim 100~{\rm pc}/v_{\rm
ff}(100{\rm pc})\sim 1 {\rm Myr}$. Here $v_{\rm ff}(100{\rm pc}) $ is the free-fall velocity
at 100 pc.

During the outburst, the accretion rate is much higher so the radiation and wind from the
central AGN become much stronger. Consequently, as shown by the plots in the middle column,
two low-density and high-temperature outflowing regions are quite evident in the polar region.
This is clearly driven by the wind. The temperature of the wind region is as high as $\sim 10^9$
K; such a high temperature is reached because the kinetic energy of the wind is converted into
the thermal energy. We can see from the figure that the wind region extends up to $\sim 100$
pc. Since this figure is a snapshot, actually the wind can reach much further away,  $\sim 20$
kpc, as we will discuss later.  Star formation in the wind region will be strongly suppressed.
Close to the equatorial plane of the galaxy, there are many high-density and low-temperature
gas clouds, which are partly formed by the squeezing due to the wind. This place is ideal
for star formation.  This gas is fueling the black hole and causes the high accretion rate of
the AGN.   The strong mechanical feedback by wind and radiative feedback by  radiation will
make the accretion rate of the  AGN strongly decrease, and thus the outburst quickly decays. The
decaying phase is shown by the right plots. We can see that again the spatial distributions of
density and temperature of the gas in the central region of the galaxy become smooth, similar
to the phase before the outburst. But different from it, in most of the region within several
hundred pc, the gas is outflowing. This is due to the AGN feedback.

The three plots in the right column correspond to $t=1.815$ Gyr (immediately before the
outburst), 1.82 Gyr (close to the peak of the outburst), and 1.83 Gyr (just after the outburst),
respectively. The minimum and maximum  accretion rates in this time interval are $ 10^{-4}
\dot{M}_{\rm Edd}$ and $10^{-2} \dot{M}_{\rm Edd}$, respectively. Before the outburst, the
spatial distribution of density and temperature are also smooth, although not as smooth as
in the left column. We see from the radial velocity plot that the gas in the central region
is outflowing. This is because winds exist in the hot accretion mode. This also explains why
the calculated accretion rate is so low although density is relatively high. With the time
elapsing,  the gas becomes cooler due to radiation, so the accretion rate increases and the
AGN enters into the outburst phase (the middle plot). Similar to the case of the left column,
we can also clearly see the two low-density and high-temperature outflowing region, which is
obviously driven by the wind in the hot mode. The difference is that now the wind region is
less obvious in the figure. This is of course  because the accretion rate is much lower so
the wind is weaker in the hot mode. Another difference between this plot and that in the left
column is that the temperature of the gas around the equatorial plane is now higher, $\sim 10^8$
K. Such a temperature is also close to the Compton temperature in the hot accretion mode (refer
to  eq. (\ref{hottemperature})). This is likely because of the Compton heating.  The decaying
phase is shown by the right plots. Compared to the cold-mode outburst in the left column, the
density and temperature also become smooth again in the central region of the galaxy. Similar
to the left column, the gas within several tens of pc is also outflowing; but in this case,
the outflowing velocity becomes smaller and the outflowing region also shrinks.

\subsection{Light Curve of AGN Luminosity}
\label{subsec:lightcurve}

\begin{figure*}
\centering
   \includegraphics[width=0.45\textwidth]{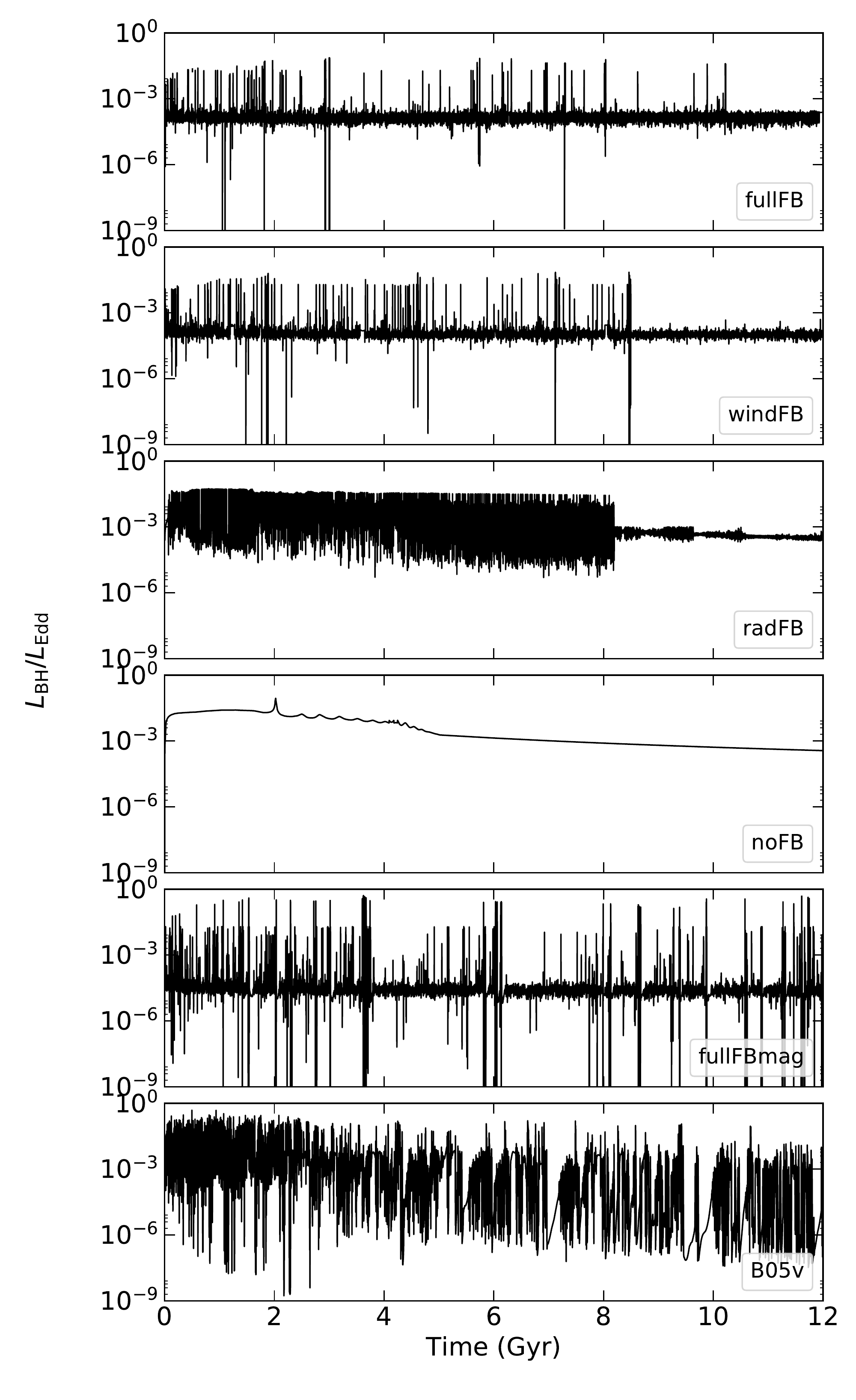}
   \includegraphics[width=0.45\textwidth]{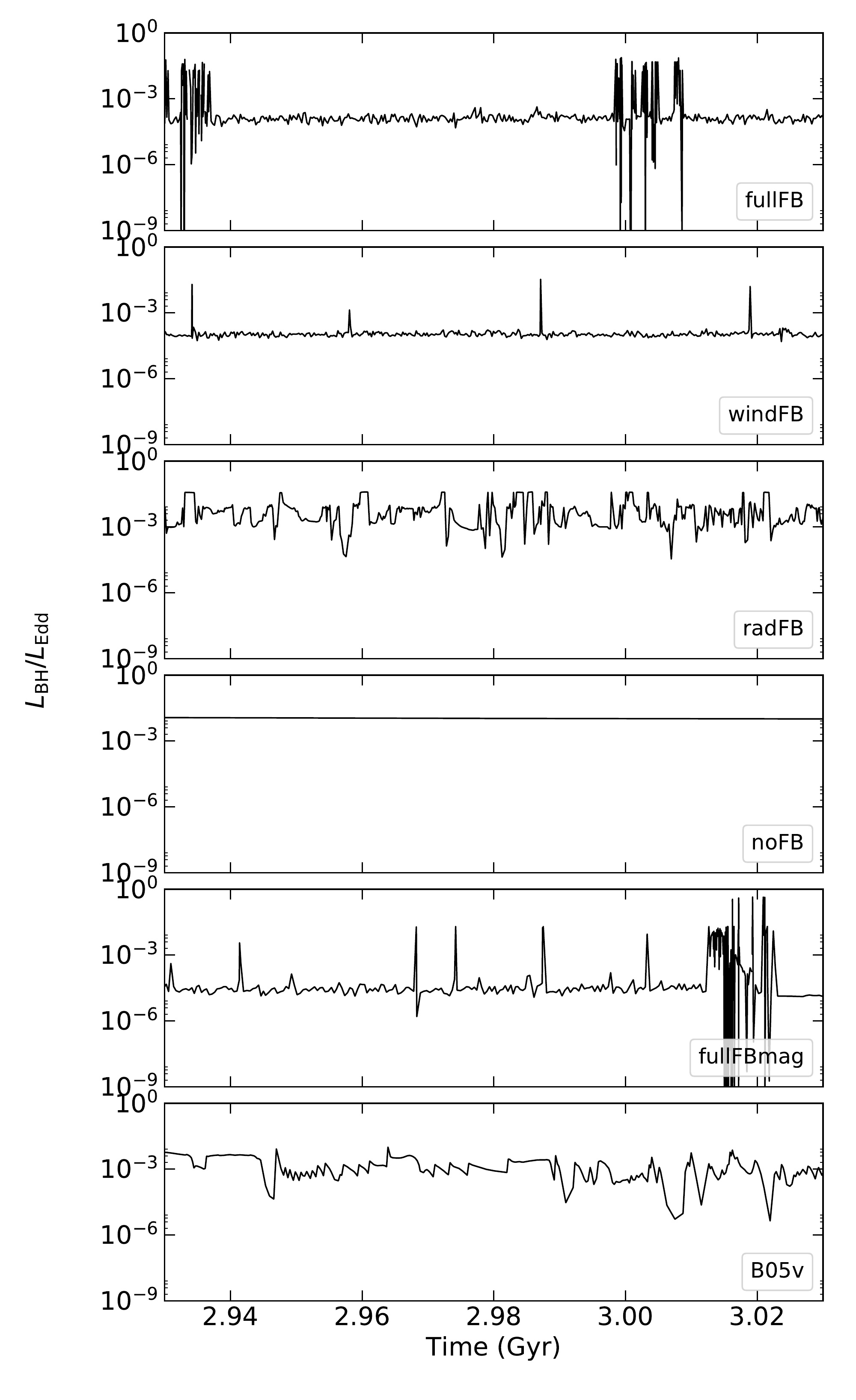}
       \vspace{0.1cm}
    \caption{Light curves of AGN luminosity as a function of time for various models.  The left
    panel is for the whole simulation time while the right panel is for  a zoom-in time made
    between 2.93 and 3.03 Gyr. The  data dump time intervals are 2.5 Myr (left panel)
    and 0.1 Myr (right panel), respectively.\\~~~~~} \label{fig:ldot}
\end{figure*}

\begin{figure}[!htbp]
    \centering
    \includegraphics[width=0.5\textwidth]{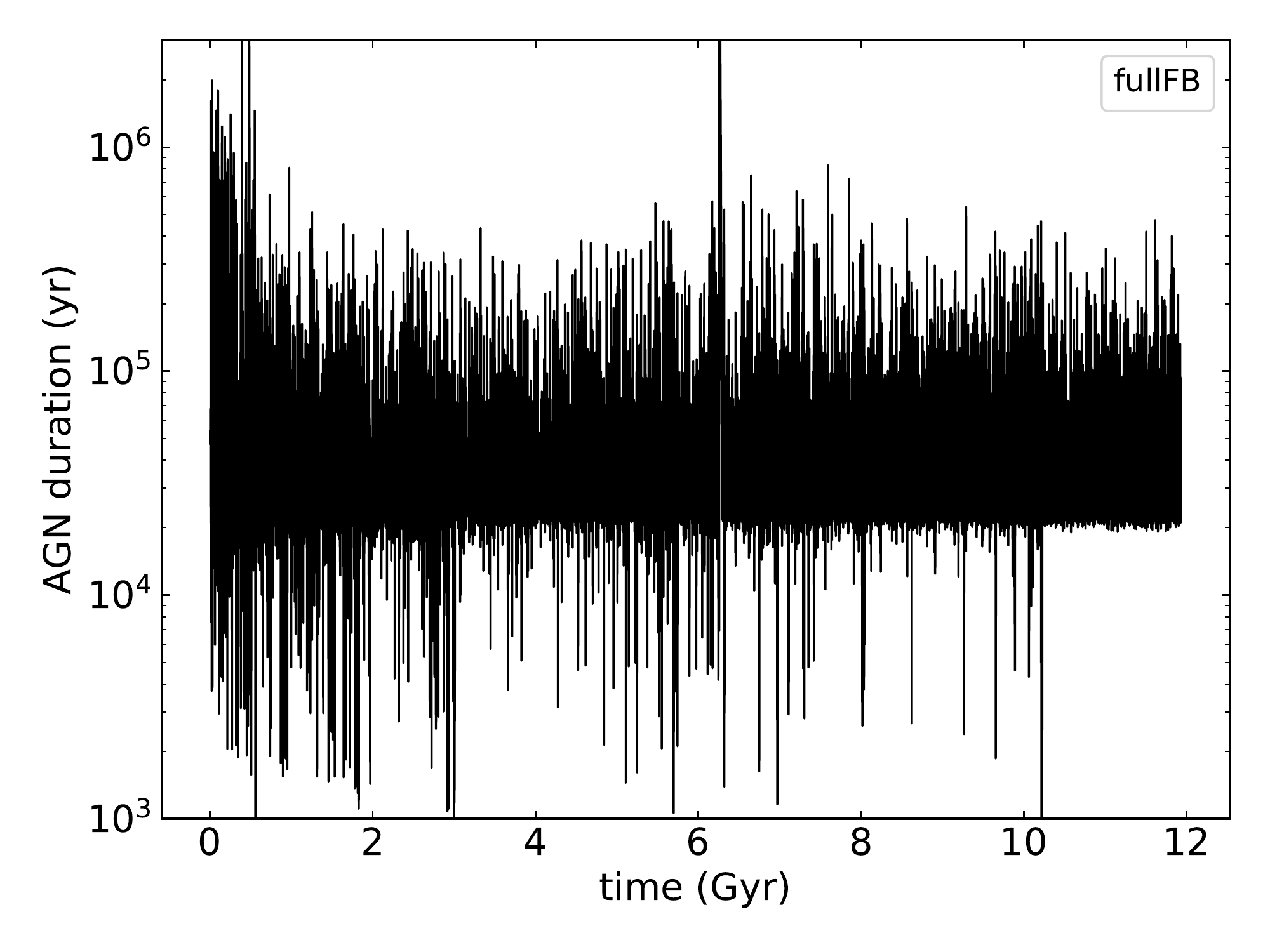}
    \caption{The duration (or lifetime) of AGN outbursts as a function of evolution time for the fullFB model. }
    \label{fig:agnlifetime}
\end{figure}

Fig.~\ref{fig:ldot} shows the light curves of the central AGN   for each model.  For comparison,
the ``fullFBmag'' and ``B05v'' models  are also included.  The left panel is for the whole
simulation time, while the right one is the zoom-in part made between 2.93 and 3.03 Gyr. Note
that for the clarity, when we draw the light curves in the left panel, we choose the data
point so that the two adjacent ones have a relatively large time interval; in this case some
outbursts are filtered out. But for the right panel of this figure we use the exact simulation
data. We see that the light curve of AGN for the noFB model is featureless; once AGN feedback
is included in the model, as we explain in the last subsection, the AGN luminosity strongly
fluctuates. From the right zoom-in panel for the fullFB model,  we can see that the AGN spends
most of its time in the low-luminosity phase, with the typical $L_{\rm BH}\sim 10^{-4}L_{\rm Edd}$. We
will discuss the AGN duty-cycle in detail in \S\ref{dutycycle}.

We can see from the figure that the variability amplitudes for the radFB and windFB models are
roughly similar, and both of them are similar to the fullFB model. This result indicates that
both the mechanical feedback by wind and the radiative feedback by radiation can cause similar
amplitude of  the AGN variability. However, there is also an important difference. In the
time-average sense, especially from the right zoom-in plot, we can see that the AGN luminosity
in the radFB model is $\sim 10^{-2} L_{\rm Edd}$, almost two orders of magnitude larger than
that in the windFB model, which is $\sim 10^{-4}L_{\rm Edd}$. The main reason for such a big
difference is  because of the difference of the ``typical length scale of feedback''. The
length scale for wind (eq. \ref{windlength}) is several orders of magnitude shorter than that
for radiation (eq. \ref{radlength}).  Therefore,  wind can efficiently deposit its momentum
and energy into the  ISM in a small volume around the black hole, and thus significantly reduce
its mass accretion rate. Radiation can only deposit its energy and momentum to the ISM within
a much larger scale, and thus is not efficient in reducing the accretion rate.  In addition,
as shown by Fig. \ref{fig:windradcomp}, in the cold mode, the momentum flux of wind is larger
than that of radiation. This means wind can more effectively push the gas away from the black
hole to reduce the accretion rate.  We can see that in the fullFB model, the ``baseline'' AGN
luminosity is very similar to that of the windFB model. This indicates that the mass accretion
rate of the black hole is controlled by the wind feedback rather than by the radiation. This is
what we expect from our analysis presented in  \S\ref{subsec:comparison}. This also explains
that the growth of the black hole mass in the windFB model will be much  smaller than that
in the radFB model, as we will discuss in \S\ref{subsec:BHgrowth}. However, by comparing the
right zoom in plots of the fullFB and windFB models, we can see that their light curves are
still different, with more outbursts in the fullFB model than in the windFB. This indicates
that feedback by radiation and wind may couple together and neither of them can be neglected.

The AGN variability amplitude in both the windFB and radFB models suddenly becomes much smaller
after $\sim 8$ Gyr. In addition, different from the epoch before 8 Gyr during which the AGN
oscillates between the cold and hot accretion modes, after 8 Gyr the AGN always stays in the
low-luminosity hot accretion mode.  From Fig. \ref{fig:agnoutburst}, we see that the outbursts
of the AGN is because of the accretion of dense gas such as cold clumps.  The radiation and
especially wind from the AGN is very helpful to the formation of such clumps, since they can
perturb the ISM and make its density distribution highly inhomogeneous.  Obviously, such kind
of perturbation is most strong in the case of cold accretion mode.  This argument also explains
why the AGN can reach luminosities as low as $\sim 10^{-9}L_{\rm Edd}$ before $\sim 8$ Gyr,
which is also because of the strong interaction between AGN and ISM.  Because the mass-loss
rate from the stellar evolution gradually decays with time and because of the mass lost in the
galaxy wind, the gas in the galaxy fueling the black hole becomes fewer with time.  This is
verified by the gradual decrease of the light curve of the noFB model.  Consequently, after
$\sim$ 8 Gyr,  the AGN can no longer reach the cold accretion mode, thus the perturbation to
the ISM becomes much weaker so the clumps are rarely formed. This explains the disappearance
of the outbursts after $\sim$ 8 Gyr in both the windFB and radFB models.  For the fulFB model,
however, we can still find a few outbursts after 8 Gyr. This is because in this model we have
both radiative and mechanical feedback thus the perturbation to the ISM is stronger compared
to the radFB and windFB models.

Now let us focus on the late epoch of the windFB and radFB models. We can see from the figure
that when $t\ga 8$ Gyrs, the AGN luminosity $L_{\rm BH}\la 10^{-3}L_{\rm Edd}$ thus AGN always stays
in the hot accretion mode for both models. We can see some variability of AGN in both light
curves. The variability amplitude in both cases is small. The main reason is that the radiation
and wind are very weak at such a low luminosity. Another reason is that the gas temperature may
be high and density low, so that the typical interaction length scale for radiation and wind
become very large so the interaction with ISM is not so efficient. The presence of variability
in the two models indicates that both wind and radiation have some feedback effects even when
they are very weak, at  least in terms of modulating the accretion rate.  But their respective
mechanism may be different.  From  Fig. \ref{fig:windradcomp}, we can see that the momentum
flux of wind is  larger than radiation but the power of radiation is  larger than  wind. So
wind feedback may play its role by momentum interaction while radiation is by energy interaction
(radiative heating). The amplitude of variability in the two models are similar, which suggests
that the importance of wind and radiation may be similar. It will be an important project to
study systematically the importance of feedback by wind and radiation in the hot mode.

Now let us compare the fullFB model with the fullFBmag model. The only difference between these
two models is the initial black hole mass. The most significant difference of their light
curves is that in the fullFBmag model, the AGN stays in the high-luminosity outburst phase
for a longer duration than in the fullFB model. This is explained as follows. Remember that
the AGN mass accretion rate is controlled by the wind feedback. When the wind is stronger,
the accretion rate is more strongly reduced  thus it is harder for the AGN to recover to
the high-luminosity outburst phase. When the black hole mass is higher, its accretion rate
will be higher thus the bolometric luminosity higher. From eqs. (\ref{eq:coldwindmass}) and
(\ref{coldwindvelocity}), both the mass flux and velocity of wind are proportional to the
bolometric luminosity. So a heavier black hole has stronger wind, thus the duration for the
AGN to stay in the high-luminosity outburst phase is shorter.

The fullFBmag and B05v models have the exactly same initial black hole mass, but the AGN physics
adopted in the two models is quite different. Such a difference produces very different AGN
light curves.  From the right zoom in plots, we can see that the typical AGN luminosity of
the fullFBmag model is $10^{-4}-10^{-5}L_{\rm Edd}$; while for the B05v model it is more
than two orders of magnitude  higher, $\sim 10^{-2}L_{\rm Edd}$. The main physical reason
for this difference is that the wind adopted in our current paper in both the cold and hot
feedback modes are much stronger than that in \citet{Gan:14}. A minor reason is that  the
high value of $T_{\rm C,hot}$ adopted in \citet{Gan:14} makes the temperature of the gas
surrounding the black hole as high as $10^9~{\rm K}$, so $l_{\rm wind}$ becomes much larger,
$\sim 5{\rm kpc}$, thus the mechanical feedback becomes much less effective. The less powerful
wind and its low feedback efficiency  in the B05v model result in a consequence that the AGN
variability  is dominated by the radiation instead of wind.  This is confirmed by the rather
similar pattern  of light curves between the radFB and B05v models (refer to the right zoom
in plots).  In contrast to the fullFBmag and fullFB models, the B05v model predicts that the
AGN will spend a high fraction of its time staying in the high-luminosity phase. This is  not
consistent with observations. This indicates the importance of having a correct AGN physics.

Another important consequence of changing the AGN physics  is the effect on the
typical AGN lifetime (or duration). The AGN lifetime for the fullFB model is shown in
Fig. \ref{fig:agnlifetime}. The typical lifetime of the fullFB model is $\sim 10^5{\rm
yr}$. We define the  ``on'' and ``off''  of the AGN by comparing its luminosity with the
baseline luminosity of the fullFB model, which is $\sim 10^{-4}L_{\rm Edd}$ according to
the zoom in plot of fullFB in Fig. \ref{fig:ldot}. As a comparison, the AGN lifetime of the
B05v models in \citet{Gan:14} is  roughly $\sim 10^7 {\rm yr}$  (refer to their Fig. 2 and
the right panel of Fig. \ref{fig:ldot}). So with the new AGN physics adopted in this paper,
the AGN lifetime becomes much shorter.  This new value is consistent with the observations
\citep[e.g.,][]{Martini:03,Keel:12,Schawinski:15}. For example, based on the time lag between
an AGN switching on and the time the AGN requires to photoionize a large fraction of the host
galaxy, \citet{Schawinski:15} estimate that the AGN typically lasts $\sim 10^5{\rm yr}$.

\subsection{Mass Growth of the Black Hole}
\label{subsec:BHgrowth}

\begin{figure*}[!htbp]
    \begin{center}$
        \begin{array}{cc}
            \includegraphics[width=0.53\textwidth]{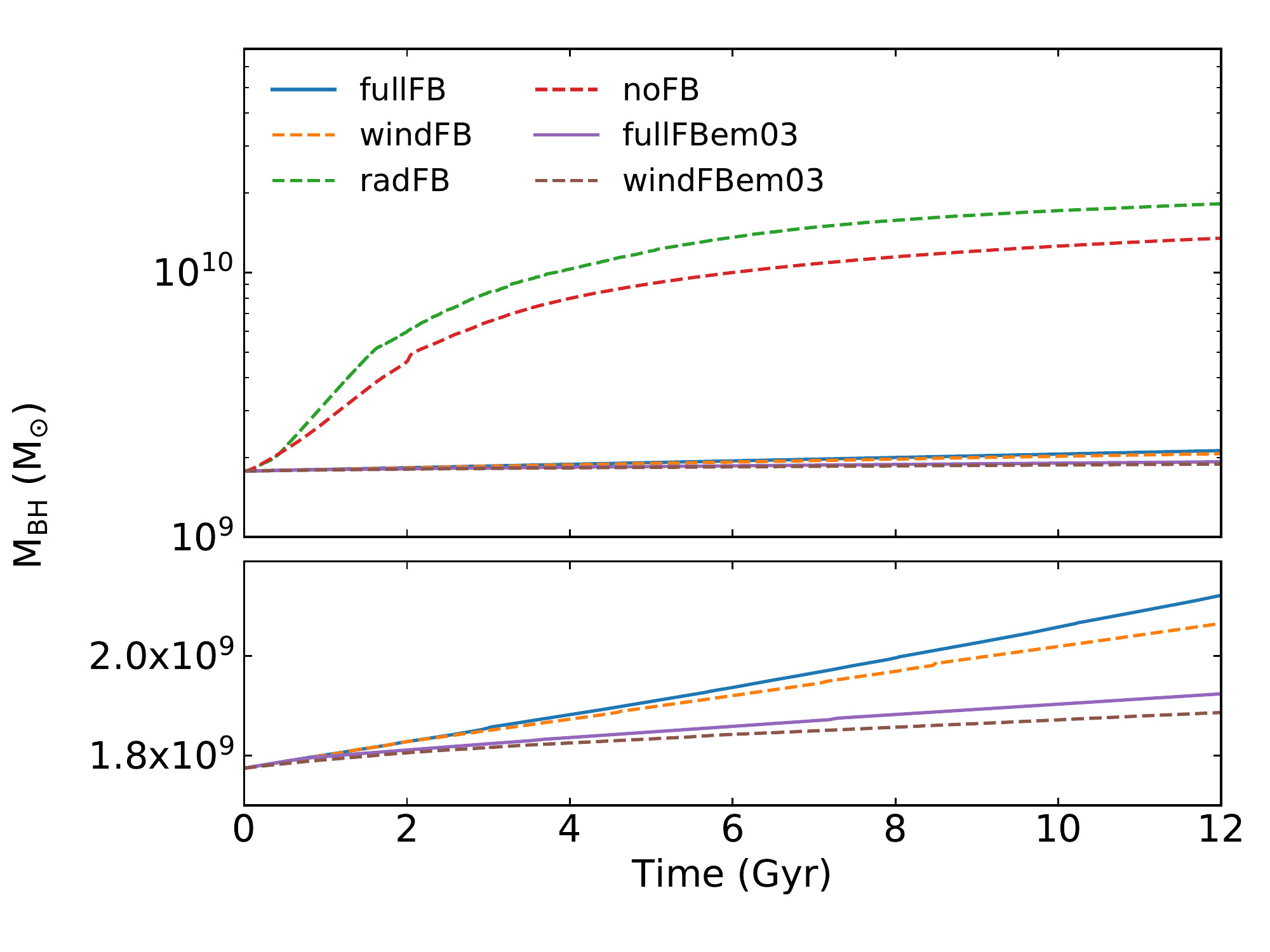} &
            \includegraphics[width=0.45\textwidth]{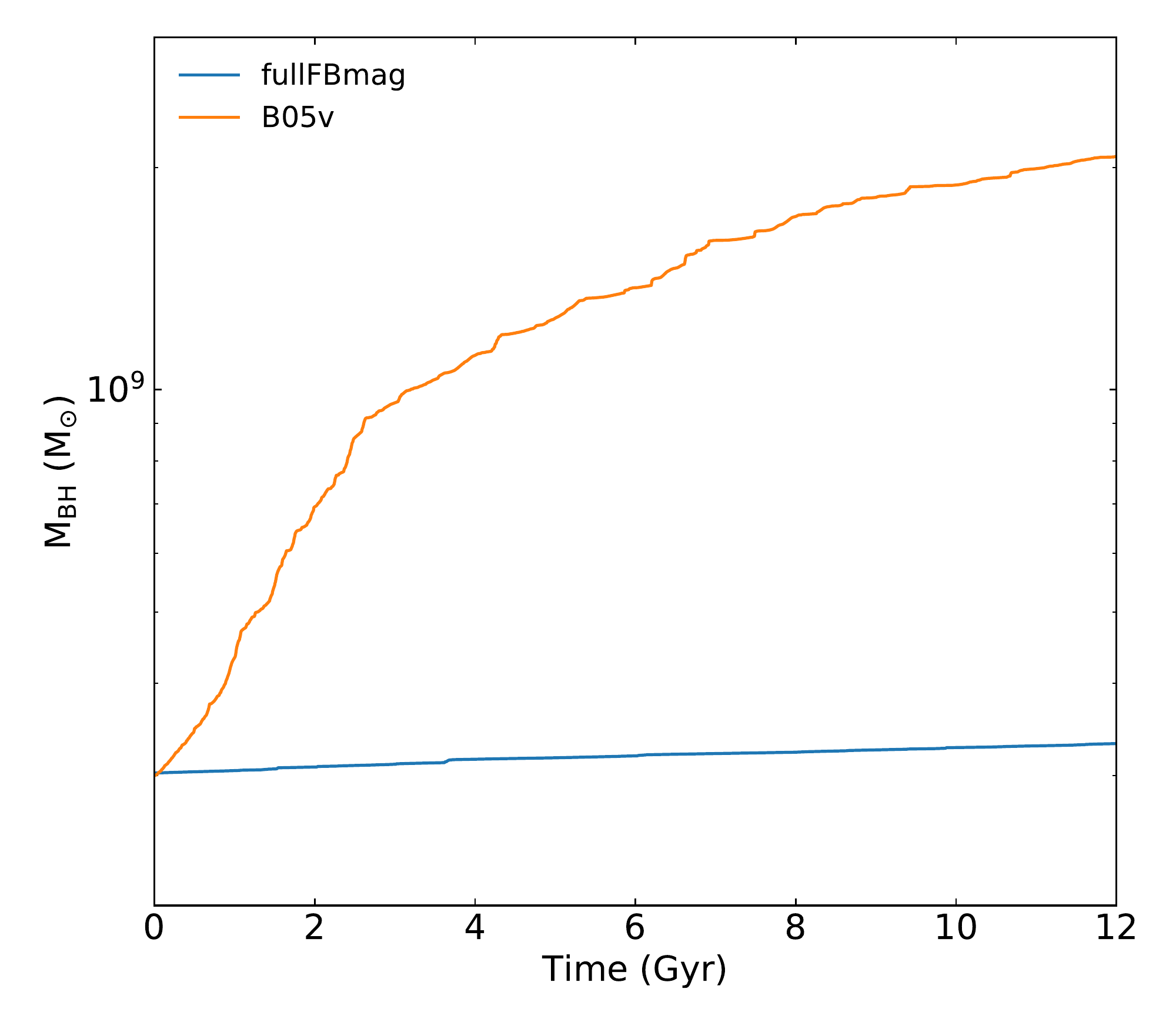}
        \end{array}$
    \end{center}
    \caption{Evolution of black hole mass for various models. In the left panel, the upper plot shows the
    black hole mass growth for the models with an initial black hole mass of $1.8\times10^{9}
    \, M_{\odot}$, and the bottom plot shows the black hole mass growth for fullFB, windFB, and
    fullFBem03 models in detail with linear scale.  The right panel is for the two models with an initial black hole mass of $3\times10^{8}\,M_{\odot}$.}
    \label{fig:bhmass}
\end{figure*}

\begin{figure}[!htbp]
    \centering
    \includegraphics[width=0.5\textwidth]{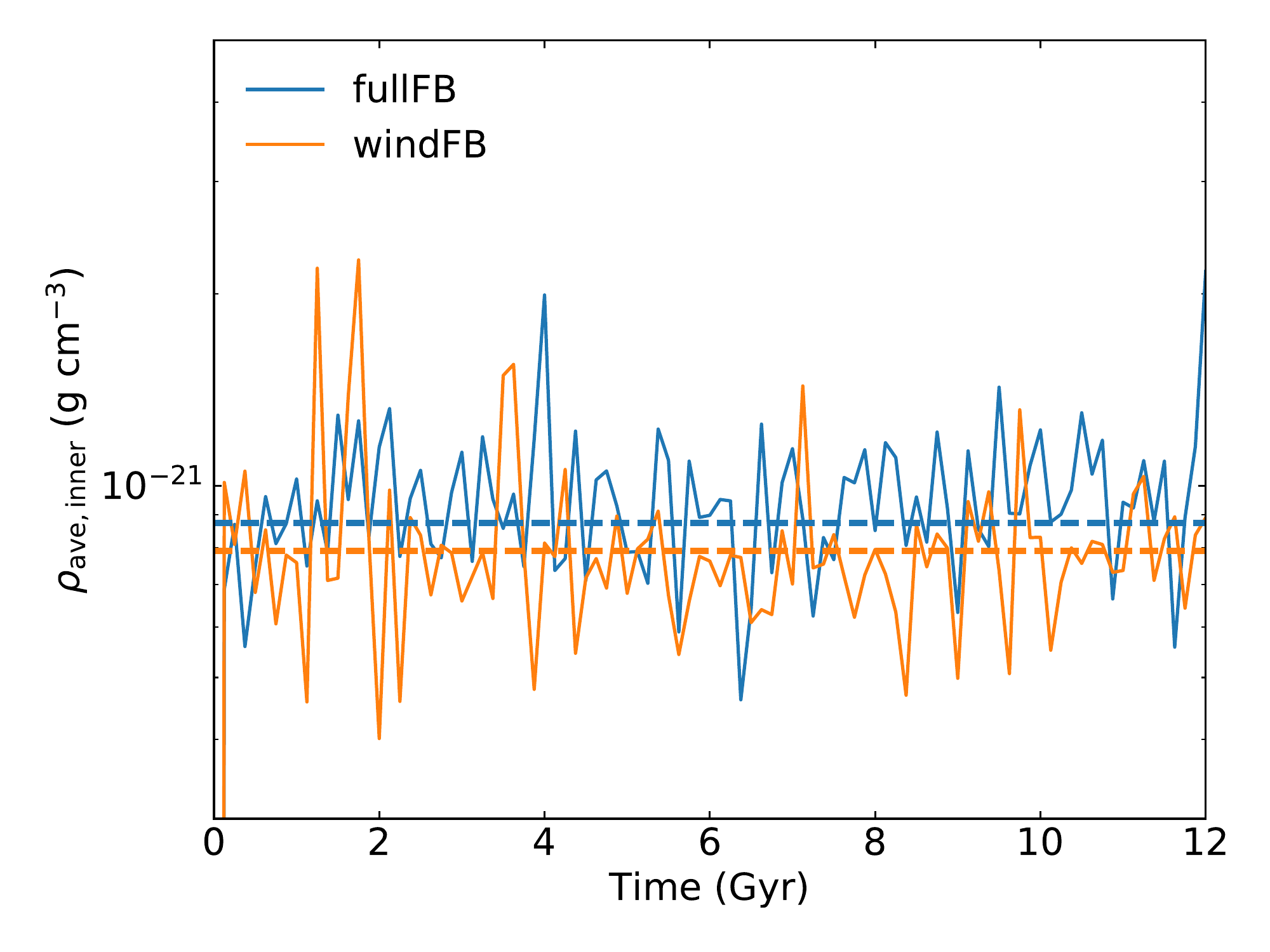}
    \caption{Time evolution of the gas density at the inner boundary of the simulation for
    two models.  The horizontal  dashed lines indicate the time-averaged density. }
    \label{fig:denin}
\end{figure}

Fig.~\ref{fig:bhmass} shows the evolution of black hole mass for each model.  Remember that the
initial mass of the black hole is set to be $M_{\rm BH,init} = 6\times 10^{-3}\, M_{\star,\rm
init}=1.8\times 10^9\msun$ \citep{Kormendy:13}.  For the noFB model, at the end of evolution the
black hole mass reaches as high as over $10^{10} \msun$, which is obviously too large compared
to observations. Since there is no AGN feedback, the gas keeps accreting onto the black hole
with little disturbance. In such a situation, mainly only star formation can deplete some gas
in the galaxy and reduce the black hole accretion rate which is not very efficient. This is
why the black hole can grow to a very large mass.

In the radFB model where only radiative feedback is considered, the black hole mass becomes even
slightly larger than the noFB case. This is surprising at the first sight, since one may think the
energy input to the gas by AGN radiation should reduce the accretion rate. While this effect must
be there, another effect seems to be more important. That is, when the AGN radiation is included,
the star formation is suppressed to some degree, mainly due to the  radiative heating to the ISM
(refer to Fig. \ref{fig:newst}).  Consequently, there will be more gas left and falling onto
the inner region of the galaxy to feed the black hole so the accretion rate is higher.

The growth of the black hole in the windFB model is substantially suppressed compared to the
noFB model. In fact, the final black hole mass in the windFB model is $2\times 10^9\msun$, only
slightly larger than its initial value. This indicates the high efficiency and dominant role
of the wind feedback in suppressing the mass accretion rate.  The reason has been explained in
detail in \S\ref{subsec:lightcurve}.  Such a result is in good agreement with \citet{Gan:14}
and \citet{Ciotti:17}.

However, it is apparently surprising to see from the figure that the black hole mass in the
fullFB model is slightly larger than that in the windFB model. One would expect that with
the inclusion of radiation, more energy is deposited in the ISM so the accretion rate in the
fullFB model should be smaller than windFB model. The reason for the smaller black hole mass
in the windFB model is exactly same with what we have just proposed above to explain why the
black hole mass is larger in the radFB model compared to noFB model. That is, when radiation is
included, star formation becomes weaker thus more gas is left to fuel the black hole.  This is
confirmed by  Fig. \ref{fig:denin}, which shows the density of the gas at the inner boundary
of the simulation for fullFB and windFB models. From this figure we can see that the average
gas density in the windFB model is smaller than that in the fullFB model. However, we would
caution readers that such a trend of black hole mass change with the feedback models may not be
a universal result. In fact, \citet{Ciotti:17} have found both positive and negative change
of black hole mass, depending on the mass of  galaxies.  This is likely because the physics
involved is highly non-linear and complicated.

In the model with higher radiative efficiency (fullFBem03), the mass growth of black hole is
further  suppressed compared with the fullFB model. We think this is not because of the stronger
radiative heating due to higher luminosity, but because of the stronger wind in the cold feedback
mode, since wind strength is proportional to the luminosity (eqs. (\ref{eq:coldwindmass})
\& (\ref{coldwindvelocity})).  To check this speculation, we have run another model named
``windFBem03'' (refer to Table 1). The mass growth of black hole for this model is shown in
Figure~\ref{fig:bhmass}.  Both ``windFB'' and ``windFBem03'' models include only wind feedback
(i.e., no radiation) and the only difference between them is the radiative efficiency. The
result shows that the final black hole mass in windFBem03 is smaller than in windFB.

The right plot of Fig. \ref{fig:bhmass} shows the growth of black hole mass for fullFBmag and
B05v models. Similar to the fullFB model, the growth of mass in the fullFBmag model is also
very small. On the other hand, the growth of black hole mass in the B05v model which has the
same initial black hole mass  with, but different AGN physics from, the fullFBmag model, is
nearly ten times larger. This indicates that the growth of black hole mass is mainly controlled
by the AGN physics instead of  the black hole mass.

An interesting question is that whether the model can explain the observed
correlation between black hole mass and the total mass of stars for elliptical galaxies
\citep[e.g.,][]{McConnell:13}. \citet{Ciotti:17} investigated this problem based on their model
and found that the model can explain the observation quite well within the uncertainties.
Since the AGN feedback physics adopted in this paper is different from \citet{Ciotti:17},
it is necessary to investigate this problem again based on our model.  We plan to pursue this
problem in our future work.

\subsection{Star Formation}

\begin{figure}[!htbp]
    \centering
    \includegraphics[width=0.4\textwidth]{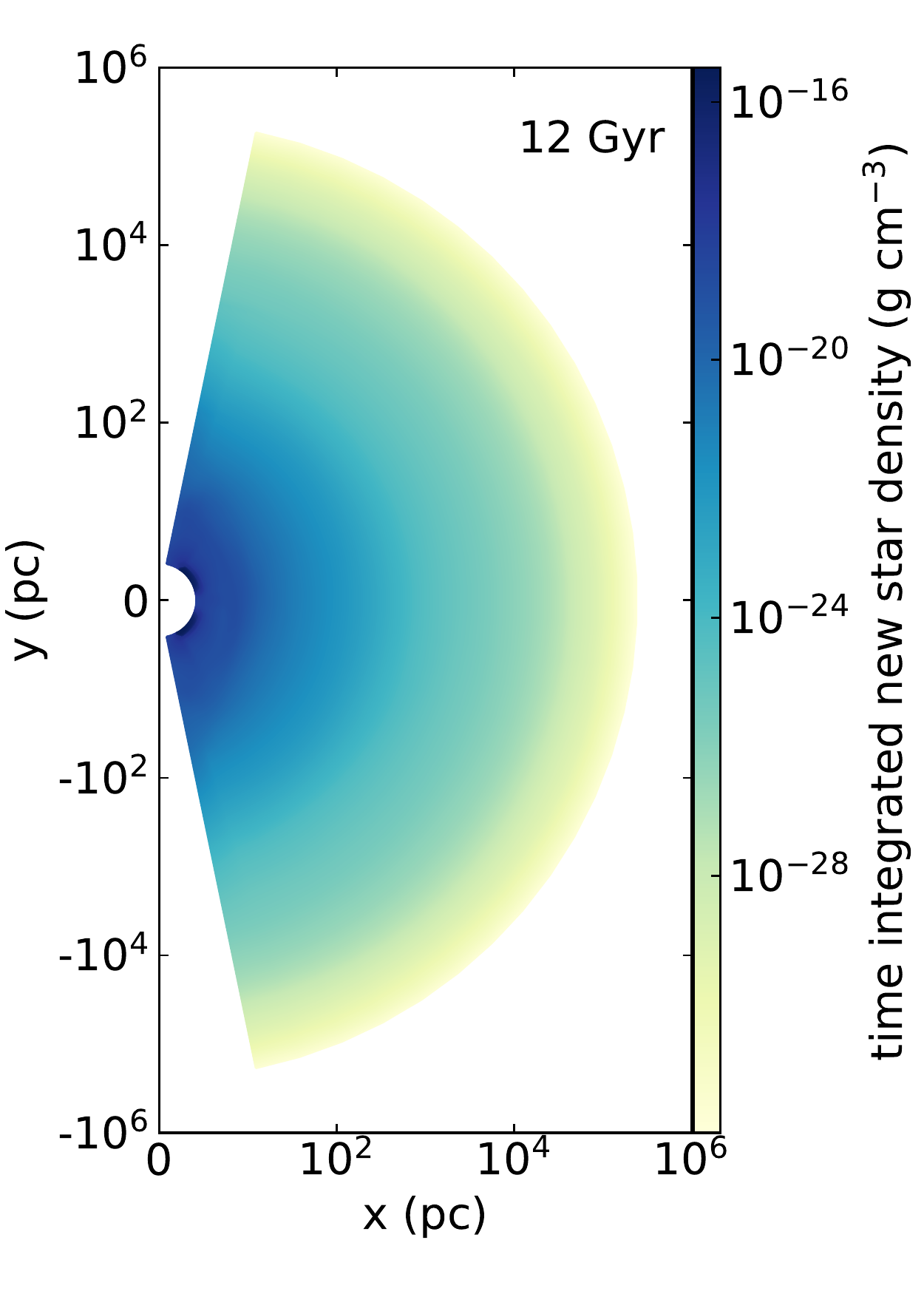}
    \caption{Time-integrated density of newly born stars at the end of the run for the fullFB model. \\~~~~}
    \label{fig:dnewst}
\end{figure}

Fig.~\ref{fig:dnewst} shows the density of newly born stars which was accumulated to the end
of our simulation.  In general, stars form massively at the cold shells or filaments. The star
formation rate is highest at the central region where density is higher, and is quite spherically
symmetric due to the low angular momentum of the gas in the galaxy.

\begin{figure*}[!htbp]
    \begin{center}$
        \begin{array}{ccc}
            \includegraphics[width=0.32\textwidth]{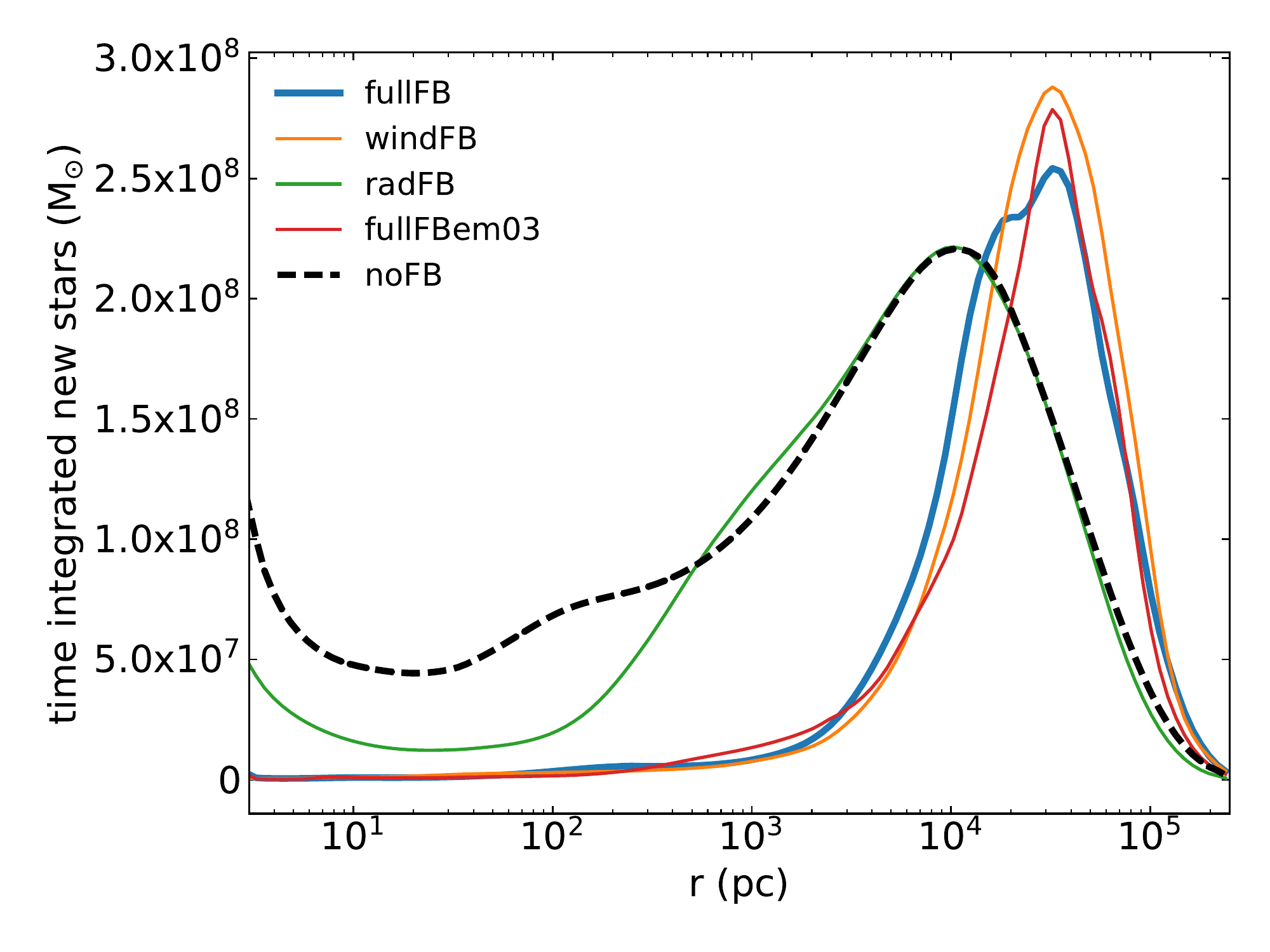} &
            \includegraphics[width=0.33\textwidth]{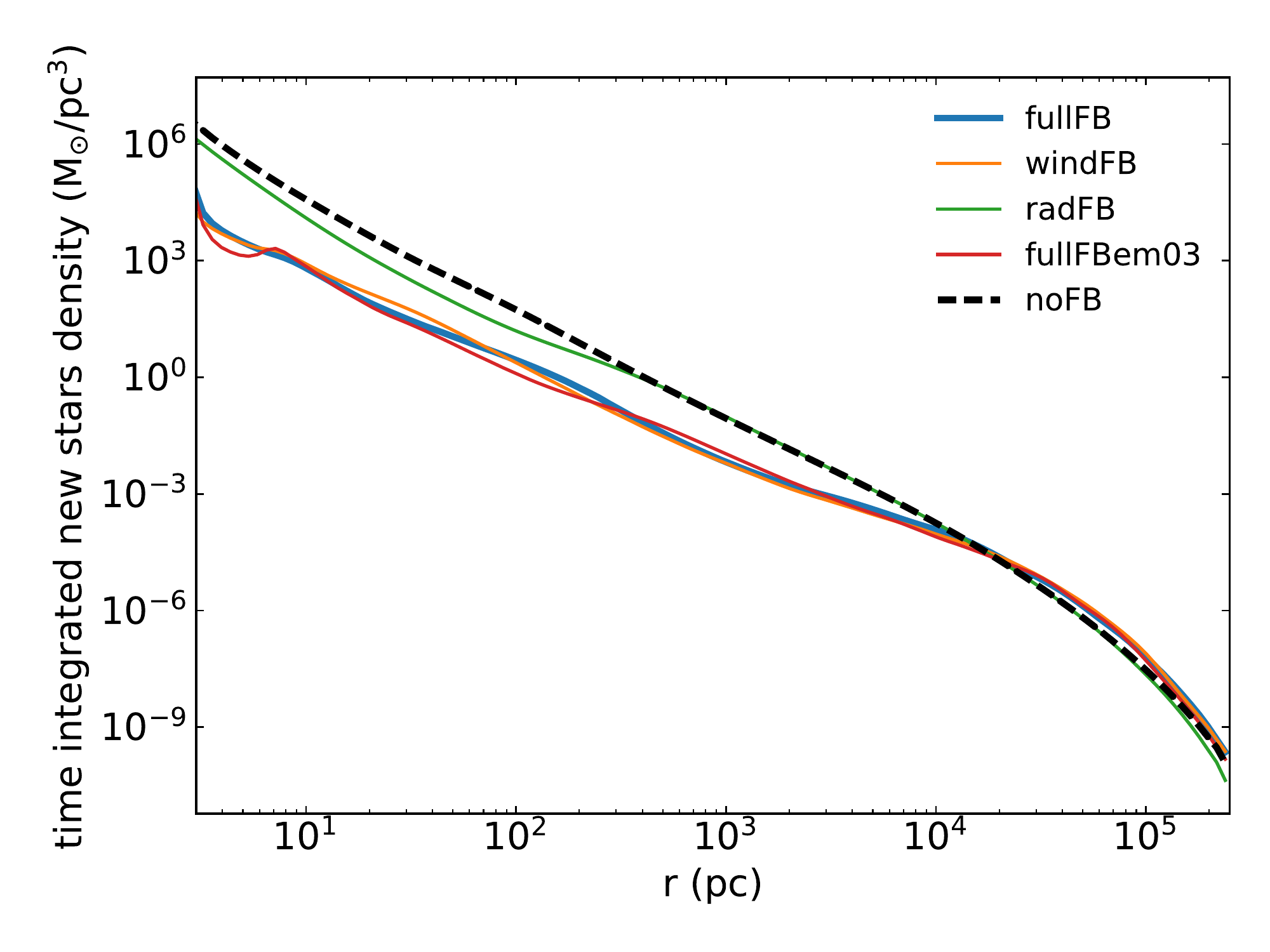} &
            \includegraphics[width=0.32\textwidth]{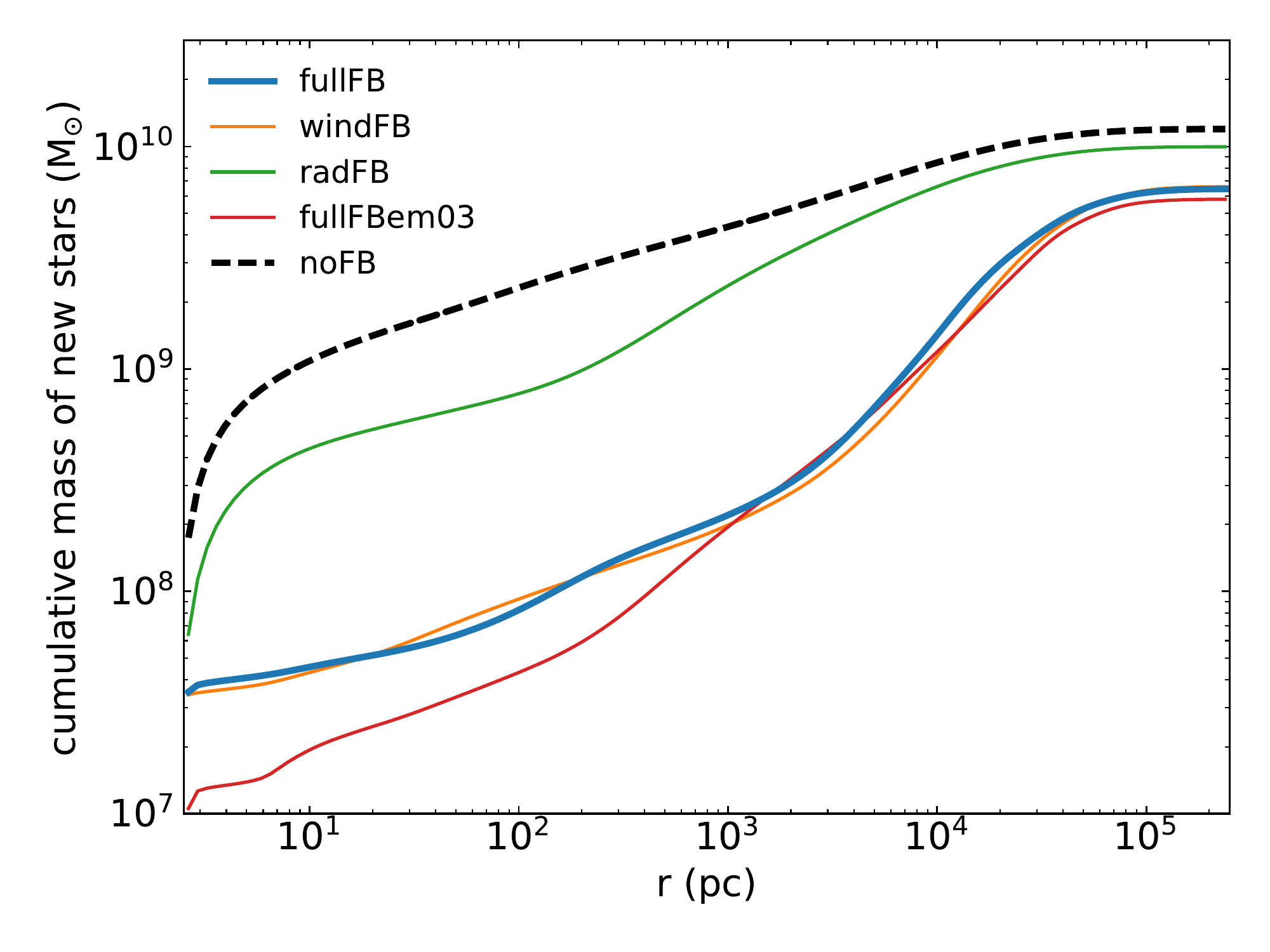}
       \end{array}$
    \end{center}
    \caption{Effects of AGN feedback on star formation for various models. Left panel:
    Time-integrated  mass of newly born stars at a given radius. Note that the peaks of each curve
    are simply due to geometry effect; see the text for details.  Middle panel: Time-integrated
    mass of newly born stars at a given radius per unit volume. Right panel: Enclosed mass of
    the newly born stars within a given radius at the end of the simulation.\\~~~~}
    \label{fig:newst}
\end{figure*}

\begin{figure}[!htbp]
    \centering
    \includegraphics[width=0.5\textwidth]{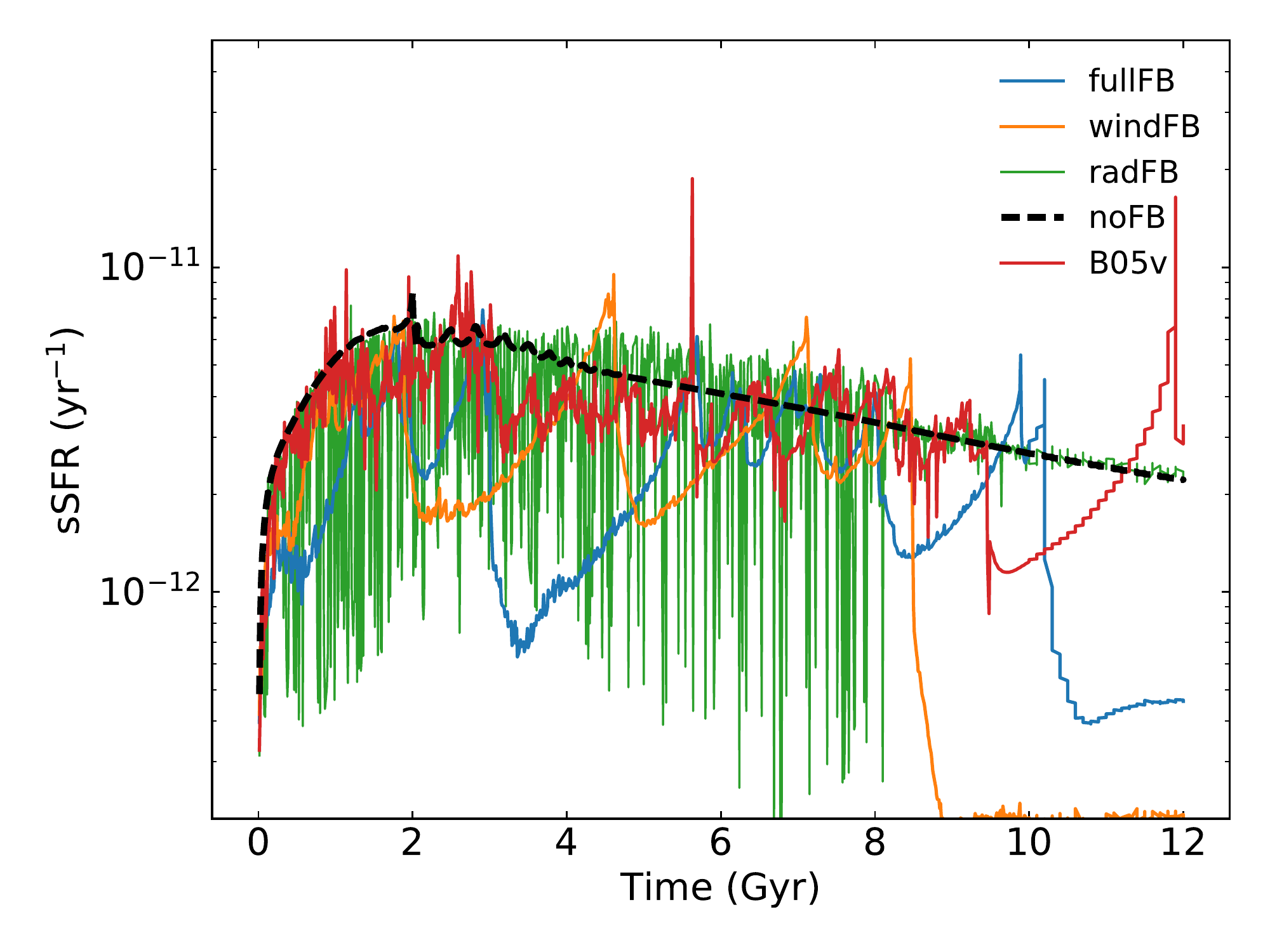}
   \caption{The specific star formation rate over time for various models. \\~~~~}
    \label{fig:sSFR}
\end{figure}

Fig.~\ref{fig:newst} shows the time- and $\theta$-integrated total mass of newly born star in our
simulation. The left panel shows the mass in each radial grid $\Delta r_i$ (note that $\Delta
r_i\propto r$) as a function of radius, while the right panel  shows the enclosed mass of the
newly born stars within a given radius. In the left panel, the peak of each curve  appears at
$r\sim 10-30$ kpc. We emphasize that this peak is only an apparent effect and does not mean
strong star formation at this radius. Its presence is because the total mass of new stars is
integrated within a radial grid $\Delta r_i$ while $\Delta r_i \propto r$, thus the volume of
each grid $\propto r^3$.  The radius of the peak, i.e., $r\sim 10-30$ kpc, corresponds to the
length scale of the galaxy $r_*$ (eq. (\ref{stellardis})) in the stellar distribution. Beyond
this radius the stellar number density sharply decreases thus there are very few gas from
stellar wind for the formation of stars.  If we normalize the mass of new stars by the volume,
the peak will disappear, as shown by the middle panel of Fig. ~\ref{fig:newst}.

Now let us see some details of the effects of AGN feedback on star formation.  The black dashed
line in the left panel denotes model noFB, its peak appears at $\sim 10$ kpc, and there is a
rapid increase at the innermost region with the decreasing radius because of the increase of
density there.  For the radFB model, denoted by the green line, it peaks almost at the same
radius with the black dashed line. At the region $r\la$ 600 pc, star formation is significantly
suppressed compared with the noFB model. This decrease is perhaps caused by the radiative heating.
This radius ($\sim$ 600 pc) is roughly equal to the typical length scale of  radiative feedback
$l_{\rm rad}$ if $\rho\sim 10^{-23} {\rm g~cm^{-3}}$ (eq.~(\ref{radlength})). Note that here we
adopt a  larger density  than that shown in Fig. \ref{fig:agnoutburst}; this is because, on the one
hand,  the density in the radFB model should be higher than that in fullFB since in radFB there is
weaker star formation and there is no AGN wind blowing gas out. On the other hand, density varies
with time while we should give more weight to the high density since star formation is easier in
that case. Beyond 600 pc, the radiation power has been used up thus radiative heating is very
weak. In this region, we can see some enhancement of the star formation compared to the noFB
mode. This is because  radiation force pushes the gas from within $\sim 600$ pc to this region.

For the windFB model,  we can see from the left plot that compared to the noFB and also radFB
models, star formation  is strongly suppressed all the way up to $r\sim 20$ kpc. The peak of
the curve also moves outward. The suppressing of star formation is obviously because of the
momentum feedback of wind, i.e., winds  push the gas away from the central region beyond $\sim
20$ kpc thus the gas density and subsequently star formation is significantly reduced. This
is consistent with the suppression of the accretion rate by wind, as we have analyzed in
\S\ref{subsec:lightcurve}. At the region $r\ga 20$ kpc, these gas  is accumulated there, so star
formation is significantly enhanced. Our simulation result that the wind can reach a distance
as far as $\sim 20$ kpc is fully consistent with the Gemini Integral Field Unit observations
to a sample of radio-quiet quasars \citep[e.g.,][]{Liu:13a}

The solid blue line in the figure denotes the fullFB model. It is quite similar to the windFB
model, which indicates that wind feedback is dominant in suppressing the star formation.
In the region of $2-20$ kpc, star formation in the fullFB model is slightly stronger than
that in the windFB model. The reason  is  because of the momentum feedback of radiation.
Radiation can push the gas away from the region within 2 kpc to this spatial range. This radius
(i.e., 2 kpc) is roughly equal to the typical length scale of  radiative feedback $l_{\rm rad}$
if $\rho\sim 10^{-24} {\rm g~cm^{-3}}$ (eq.~(\ref{radlength})).  Here $10^{-24} {\rm g~cm^{-3}}$
is assumed to be the typical density of ISM. It is lower than that in the fullFB since because
both wind and radiation are present in the fullFB model.

Comparing the fullFB model with the noFB model, we can see from the left plot of
Fig. \ref{fig:newst} that, star formation is suppressed in the inner region of the galaxy, when
$r\la 15$ kpc. This is usually called the ``negative feedback''. But at the outer region, when
$r\ga 15$ kpc, star formation is enhanced. This is the so-called ``positive feedback''.  We note
that the positive feedback effect is consistent with the theoretical result by \citet{Liu:13b} and
observations by \citet{Cresci:15}, which argued that SF can be enhanced locally at the ahead of
the AGN wind, where gas is compressed. Although whether the AGN feedback is positive or negative
depends on the location of the galaxy, the overall effect of AGN feedback on star formation
in the whole galaxy is negative, as shown by the right plot of Fig. (\ref{fig:newst}). This
result is different from \citet{Ciotti:17} where they find the total effect on star formation
is positive. The discrepancy must be attributed to the difference of the AGN physics adopted
in the two works.


In literature, many observational papers try to study the AGN feedback by investigating the
correlation between star formation and AGN activity (see reviews by \citealt{Harrison:17} and
\citealt{Xue:17}). From our simulation, we see that the  effect of AGN feedback on star formation
is very complicated. As we state above, in the time-averaged sense, the effect can be positive
or negative,  depending on the location of the galaxy. In addition to the spatial complication,
we also have temporal complications.  AGN activity varies on the timescale of $\tau_{\rm AGN}
\lesssim 1$ Myr), which is orders of magnitude shorter than the typical timescale of star
formation episodes ($\tau_{\rm SF} \gtrsim 100$ Myrs) \citep{Harrison:17}. For the whole galaxy,
star formation sometimes is enhanced but sometimes suppressed.

To illustrate this point, Fig.~\ref{fig:sSFR} shows the specific star formation rate (sSFR) over
the galactic evolution, i.e. star formation rate normalized by the stellar mass of the galaxy,
which is a quantity widely used in literature. We can see from the figure that the timescale
of variability of sSFR is much longer than the variability of AGN, as we explain above. In
general, the sSFR in the models with AGN feedback is suppressed compared to noFB model, but
occasionally sSFR with AGN feedback can also be enhanced. Another important feature we can see
is that the curve of the windFB model is not synchronous with that of fullFB model. While their
general patterns are similar, there is an obvious offset between them and the ``amplitudes''
of the fullFB model are also larger.  This indicates that, although from the time-integrated
sense wind seems to be much more important than radiation in suppressing star formation, as
we see from Fig. \ref{fig:newst}, radiation definitely also plays a very important role. The
wind and radiation likely couple together in affecting star formation.  At last, we note that
for the windFB and fullFB models, the sSFR sharply decreases at $t\ga 8$ Gyr and $t\ga$ 10 Gyr,
respectively. Such sharp decreases correspond to the sudden change of the amplitudes of AGN light
curves at the same time  for the two models shown in Fig. \ref{fig:ldot}. As we have explained
in that section, in that case, there are very few high-density low-temperature clumps in the ISM,
and thus star formation becomes very weak.

\subsection{AGN Duty Cycle}
\label{dutycycle}

\begin{figure*}[!htbp]
    \begin{center}$
        \begin{array}{cc}
            \includegraphics[width=0.5\textwidth]{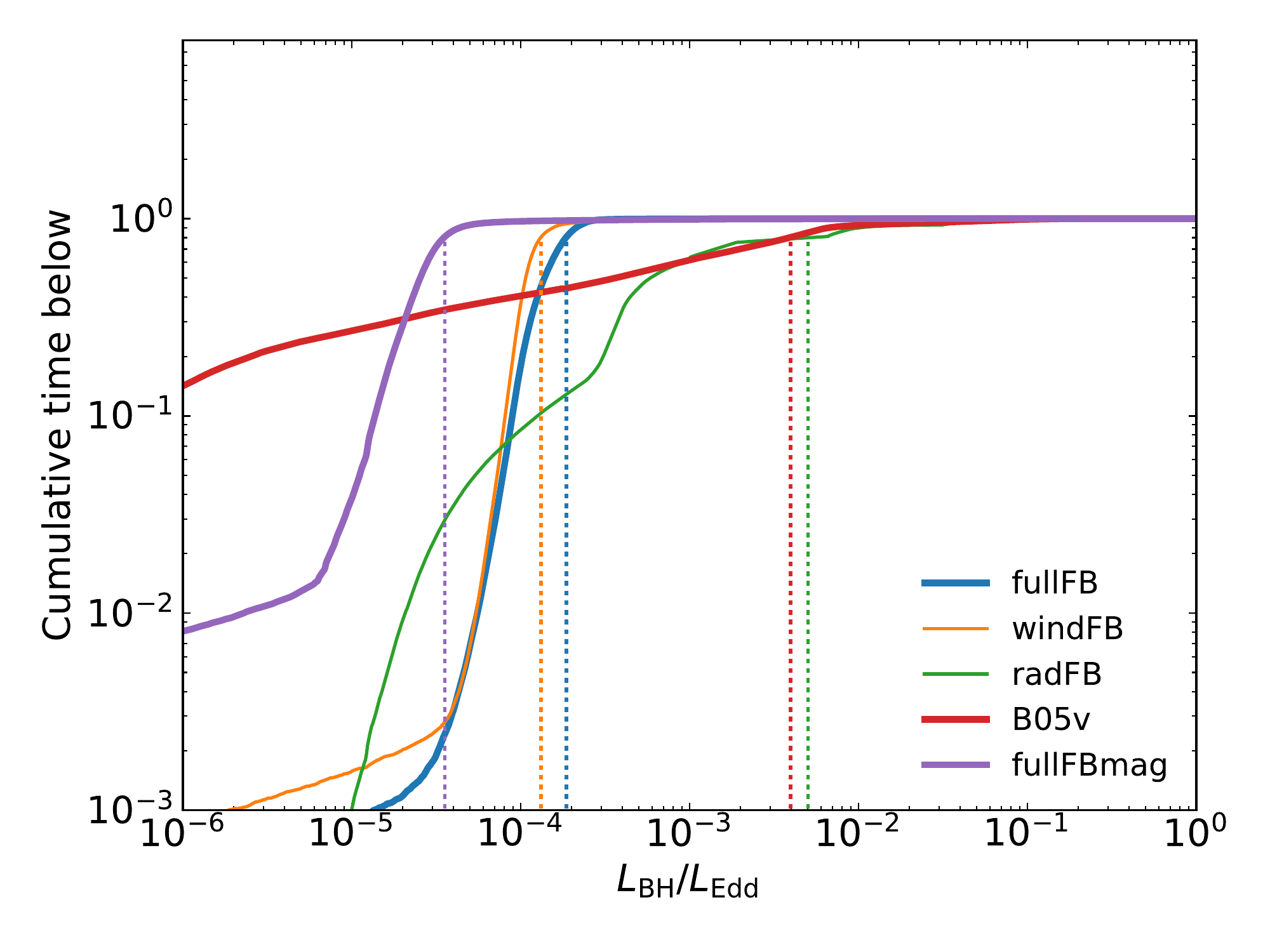} &
            \includegraphics[width=0.5\textwidth]{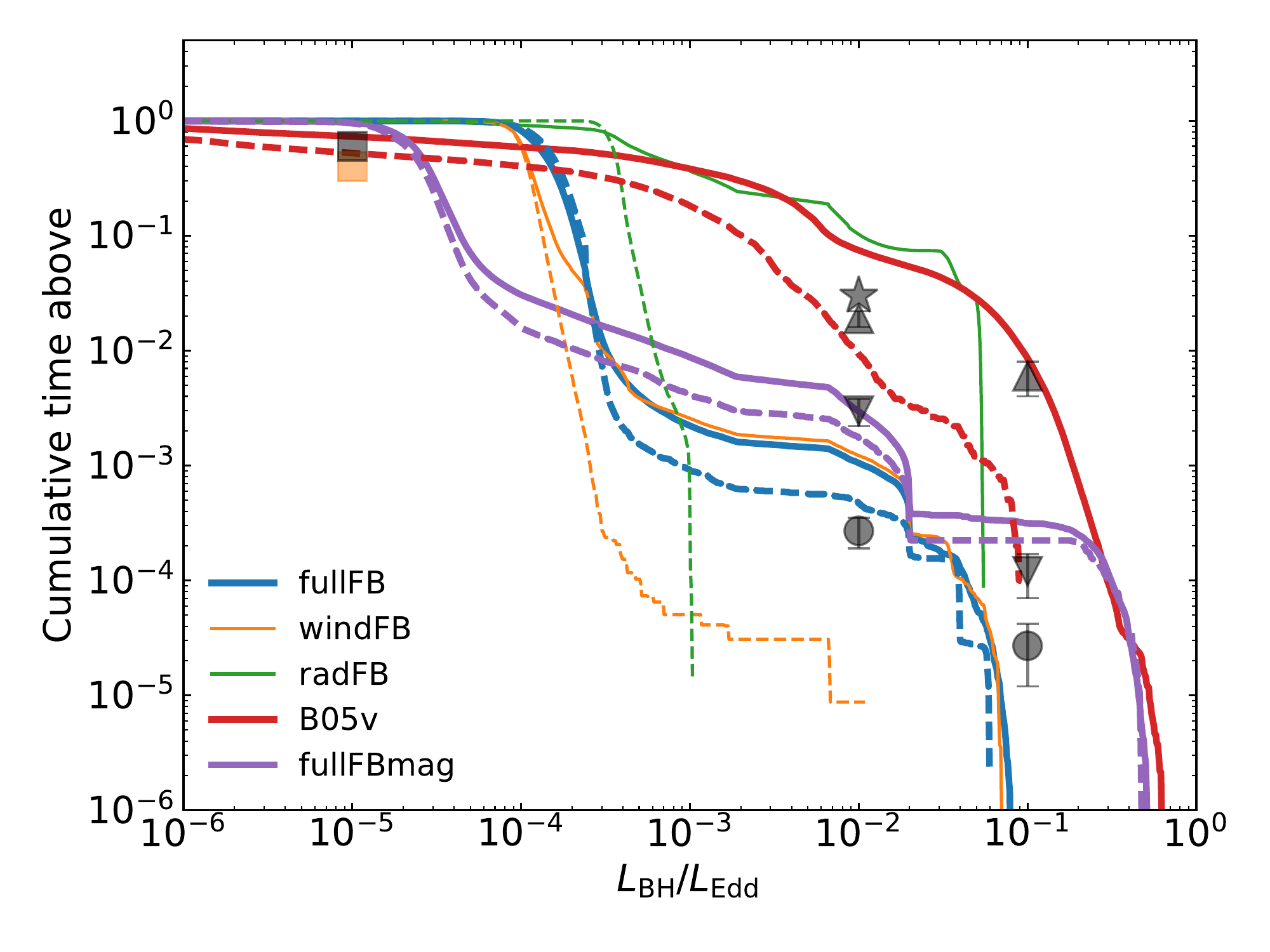}
        \end{array}$
    \end{center}
    \caption{Percentage of the total simulation time spent below (left) and above (right) the
    values of the Eddington ratio of the central AGN for various models.  In the left panel,
    vertical dotted lines indicate the Eddington ratio below which each model spends 80\% of the
    total time.  The solid lines are the values for the entire time, and the dashed lines in the
    right panel are the values for the last  2 Gyrs, which can be compared to observations [square:
    \citet{Ho:09}; circle: \citet{Greene:07}; upward-pointing triangles: \citet{Kauffmann:09};
    downward-pointing triangles: \citet{Heckman:04};  star: \citet{Steidel:03}.].\\ ~~~~}
    \label{fig:dutycycle}
\end{figure*}

\begin{figure}[!htbp]
    \centering
    \includegraphics[width=0.5\textwidth]{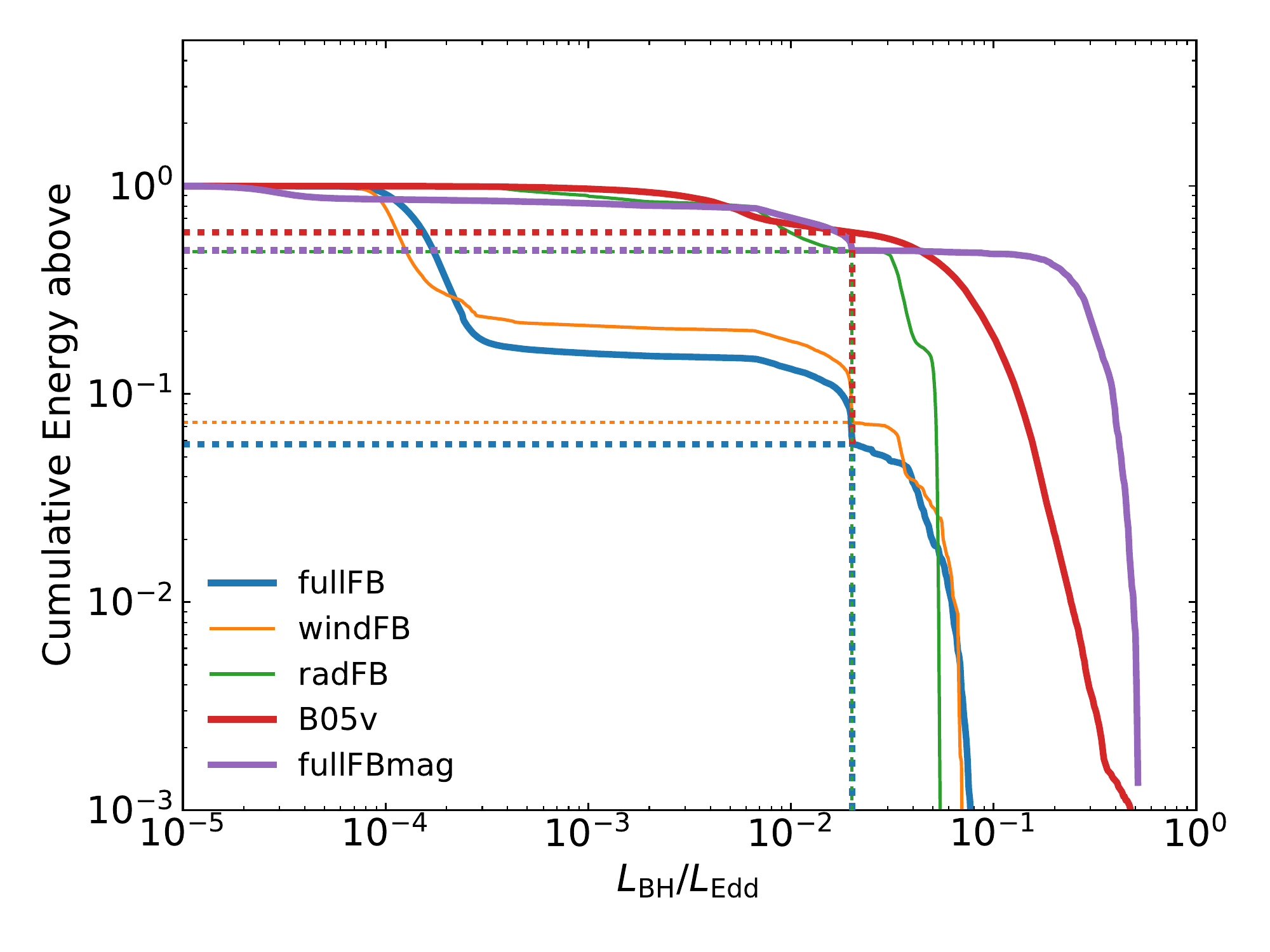}
    \caption{Percentage of the total energy emitted above the values of the Eddington ratios. The
    horizontal dotted lines represent the portion of emitted energy above the Eddington ratio
    of 0.02. \\~~~~} \label{fig:duty_L}
\end{figure}

Fig.~\ref{fig:dutycycle} shows the percentage of the total simulation time spent below (left panel)
and above (right panel) a given Eddington ratio.  In the left panel, the vertical dotted lines
indicate the Eddington ratio below which the AGN spends 80\% of its time.  Comparing various
models, we can see that from radFB to fullFB (windFB is similar), with the increase of AGN
feedback ``strength'', this Eddington ratio becomes smaller.  Among all models, the ratio is
the smallest for the fullFBmag model. This is consistent with that shown in the right panel of
Fig. \ref{fig:ldot} in which the fullFBmag model has the lowest typical Eddington ratio. For
the fullFB model, the AGN spends over 80\% of its evolution time with Eddington ratios below
$2 \times 10^{-4}$ and it spends most of its time in the hot accretion (feedback) mode. This
results suggests that potential importance of feedback effects by low-luminosity AGNs.

In the right panel of Fig.~\ref{fig:dutycycle}, we compare the simulation results with
observations.  The solid and dashed lines represent the time spent above the given Eddington
ratios for entire evolution time and for the last 2 Gyrs, respectively. The observational
data points are compiled from low-redshift sources, which are suitable to be compared with the
dashed lines. For the blue dashed line, which denotes the fullFB model  in the late epoch, the
AGN spends very little of its time being in the luminosity of $L_{\rm BH}/L_{\rm Edd}\ga 1\%$.
This can also be roughly seen from the left panel of Fig. \ref{fig:ldot}. This result is consistent
with observations \citep[e.g.,][]{Steidel:03, Heckman:04, Greene:07, Ho:09, Kauffmann:09}.

It is believed that AGNs spend  most of their time in the low-luminosity AGN phase, but emit most
of their energy during their high-luminosity AGN phase \citep{Soltan:82, Yu:02, Kollmeier:06}. To
examine this issue with our simulations, Fig.~\ref{fig:duty_L} shows the percentage of the
total energy emitted above the values of the Eddington ratios for various models.  Specifically,
the horizontal dotted lines mark the fraction of emitted energy with the Eddington ratio above
0.02. For the  fullFB models, the AGN emits $6\%$ of the entire energy at the Eddington ratio
above 0.02.  The percentage for the fullFB  model is much lower than what observations seem to
suggest. There are two main reasons for this discrepancy. One is that our simulations begin from
2 Gyr. Before 2 Gyr, the activity of AGN should be much stronger. Another reason is that the
present paper only focuses on an isolated galaxy. The percentage of energy emitted is expected
to increase significantly if we also include the cosmic accretion of cold gas from outside of
the galaxy.

We note that for the fullFBmag model,  AGN emits $50\%$ of the entire energy at the Eddington
ratio above 0.02. This is much closer to the observation compared to model fullFB. There are two
reasons for such a discrepancy between ``fullFB'' and ``fullFBmag''. One is that when the black
hole mass is smaller as in fullFBmag, the AGN becomes weaker thus it becomes more difficult to push
the gas surrounding the black hole away so the mass accretion rate is in general larger. Another
reason is that the critical luminosity (eq. (\ref{criticall})) separating the hot and cold modes
is proportional to the black hole mass. When the black hole mass is smaller, it is easier for
the AGN to switch from the hot to the cold accretion modes. These two reasons make the AGN with
a smaller black hole mass spend more time above a given Eddington ratio, as shown by the right
plot of Fig. \ref{fig:dutycycle}, and subsequently emit more energy above a given Eddington ratio.

The two models fullFBmag and B05v have the same black hole mass, we can see they
are similar in Fig. \ref{fig:duty_L}. However, these two models are quite different
in Fig. \ref{fig:dutycycle}. The fullFBmag model spends much fewer time above a given
Eddington ratio compared to model B05v. Such a discrepancy is caused by the difference of AGN
physics. Specifically, the wind in fullFBmag model is much stronger than in B05v thus the black
hole accretion rate in fullFBmag is generally much smaller.

\subsection{X-ray Properties of the Gas}

\begin{figure}[!htbp]
    \centering
    \includegraphics[width=0.5\textwidth]{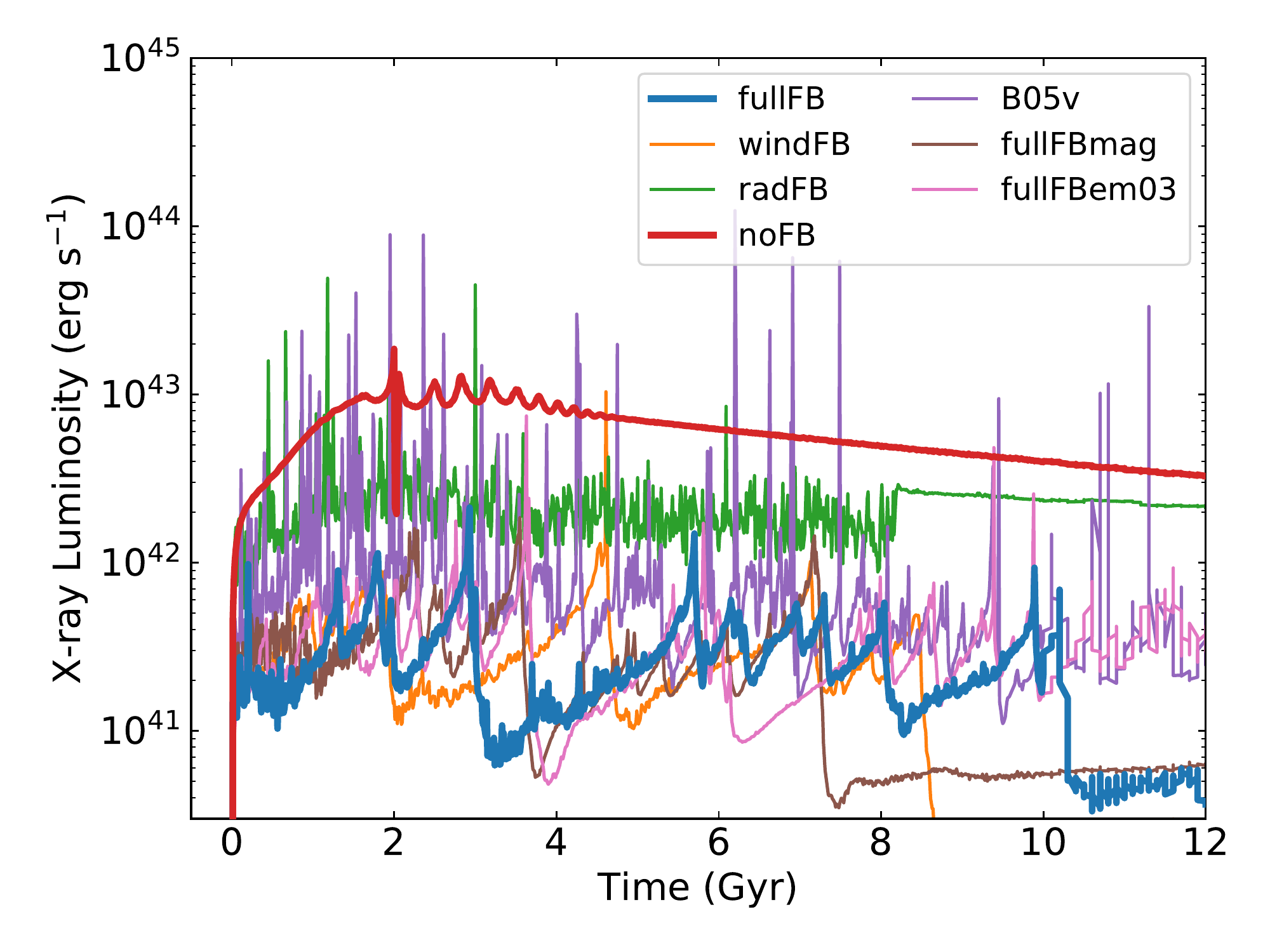}
    \caption{The evolution of the X-ray luminosity of the galaxy in the 0.3-8 keV band for various models.\\~~~~}
    \label{fig:xlum}
\end{figure}

\begin{figure}[!htbp]
    \centering
    \includegraphics[width=0.5\textwidth]{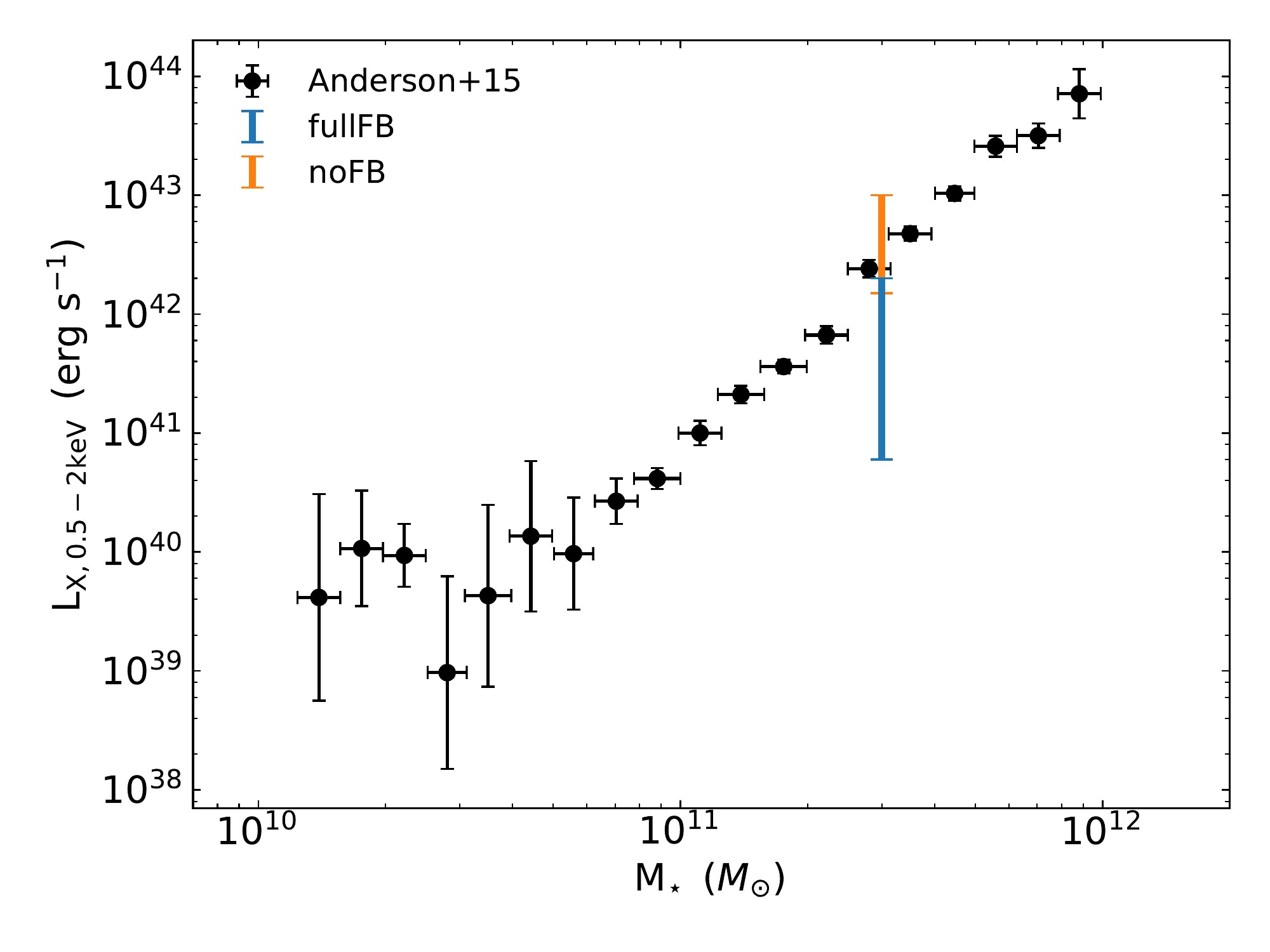}
    \caption{Comparison of the X-ray luminosity of the hot gas in the 0.5-2 keV band between
    our simulations and the observations by \citet{Anderson:15}. The black dots with error
    bars are the observational data, while the orange and blue segments denote the simulation results
    of the  noFB and fullFB models.}
    \label{fig:xlumcomp}
\end{figure}

\begin{figure}[!htbp]
    \centering
    \includegraphics[width=0.5\textwidth]{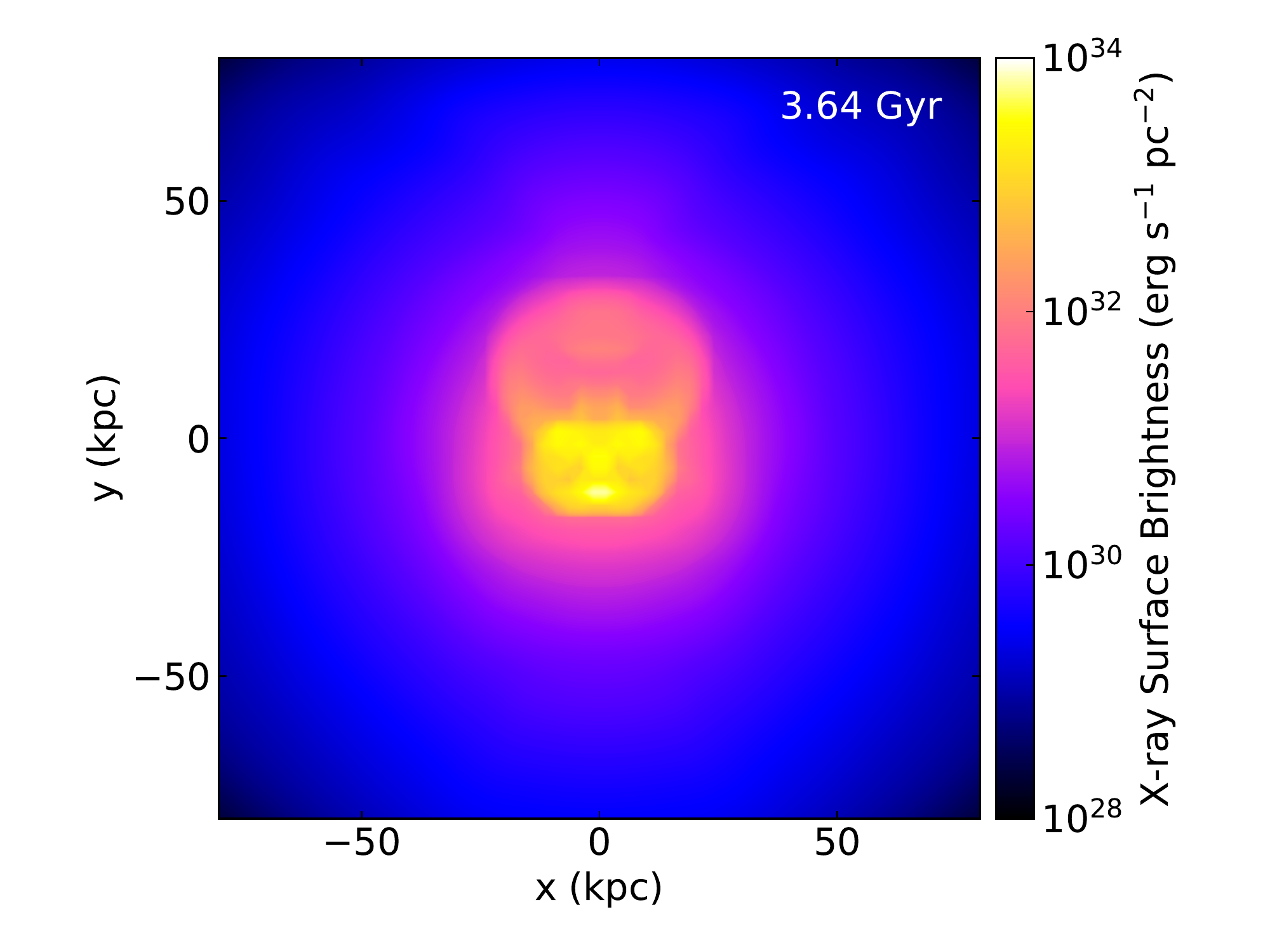}
    \caption{X-ray surface brightness in the 0.3-8 keV band for the fullFB model.}
    \label{fig:xsurf}
\end{figure}

The X-ray luminosity of the galactic hot gas content, which is mainly produced by bremsstrahlung,
is observable for active galaxies \citep[and references therein]{Brandt:15}. It is
calculated over the energy range of 0.3-8 keV (the {\it Chandra} sensitive band), which is
\begin{equation}
    L_{\rm X} = 4\,\pi \int^{\infty}_{0} \, \varepsilon \left( r \right) \,r^{2}\,dr,
\end{equation}
where the emissivity  is given by $\varepsilon (r)=n_{e}(r)n_{\rm H}(r)\Lambda \left[T
(r) \right]$, with $n_{e}$ and $n_{\rm H}$ are the number densities of electron and
hydrogen, and $\Lambda \left( T \right)$ is the cooling function. We fix the metallicity
to the solar abundance. The cooling function is calculated by the spectral fitting package
XSPEC\footnote{\url{http://heasarc.gsfc.nasa.gov/docs/xanadu/xspec/}} (spectral model APEC)
with the assumption of the collisional ionization equilibrium \citep{Smith:01},
 and the volume integrals are performed over the whole computational mesh.

Fig.~\ref{fig:xlum} shows the evolution of the X-ray luminosity, $L_{\rm X}$, for various
models. For the models with AGN feedback,  $L_{\rm X}$ oscillates in phase with the nuclear
luminosity, which is consistent with \citet{Pellegrini:12}: radiative and mechanical feedback
change the density and temperature of the gas in the galactic center, where most of $L_{\rm
X}$ is emitted.  On the contrary, in the noFB model,  the light curve of $L_{\rm X}$ is quite
smooth and  monotonously decreases as a consequence of gas depletion due to star formation.
The average value of the X-ray luminosity for various models follows this order: noFB $>$ radFB
$>$ fullFB (windFB) $>$ fullFBmag. This sequence is consistent with the light curves shown for
various models in Fig. \ref{fig:ldot}. It is interesting to note that the pattern or the shape
of the X-ray luminosity for the windFB and fullFB models is very similar to the sSFR shown in
Fig. \ref{fig:sSFR}. This is reasonable since both the sSFR and the X-ray luminosity depend on
the local properties of gas on large scales.

\citet{Anderson:15} have stacked X-ray emission from {\it ROSAT} All-Sky Survey in the X-ray
band range of 0.5-2 keV, and showed the power-law relationship between the X-ray luminosity
and the stellar mass of the central galaxy.  To compare with this observation, we have
calculated the  X-ray luminosity of the hot gas in the 0.5-2 keV band. The result is shown in
Fig. \ref{fig:xlumcomp}. Since the X-ray luminosity is variable, we show in the figure the
full range of luminosity.  We can see from the figure that the noFB model predicts a higher
luminosity than the fullFB model, as we expect, and both of their predictions  are marginally
consistent with the observations.  However, considering that our current work focuses only
on isolated galaxy and has not taken into account the gas supply from outside of the galaxy
while the X-ray luminosity values taken from \citet{Anderson:15} are for central galaxies,
the fullFB model is more promising to be consistent with observations while the noFB model
will likely over-predict the luminosity once we more properly consider the additional external
gas supply. As comparison, \citet{Eisenreich:17} recently have also compared their numerical
simulation result with the observations of \citet{Anderson:15}. They find that the simulated
value is a factor of few too high. They speculated that the reason may be because of their
initial condition of simulation. But another possible reason is the AGN physics they adopt.

In order to calculate X-ray surface brightness, we first generate three-dimensional data based
on our two-dimensional numerical simulation data by assuming axisymmetry and then integrating the
emissivity along the line-of-sight.  Fig.~\ref{fig:xsurf} shows the X-ray surface brightness for
the model fullFB during an AGN outburst.  We can see from the figure that an X-ray cavity forms
when AGN feedback expels and heats the gas in the galactic center, and such a cavity looks very
similar to those observed (e.g., \citealt{Fabian:12} and references therein).  We have also checked
the windFB and radFB models and found that the cavity is formed almost fully by the interaction
between  wind and ISM, not because of radiative heating to the ISM. We note that in literature
people usually think that the observed X-ray cavity is formed by the interaction between jet and
ISM \citep{Guo:16,Guo:18}.  The reason may be  because jets are very common but winds are harder to
be detected. However, we would like to emphasize that non-detection does not mean non-existence.
We now know that  jet is an indispensable ingredient of  hot accretion flows \citep{Yuan:14};
while in hot accretion flows we now have compelling evidences for the existence of wind, from
both theoretical \citep{Yuan:12a,Narayan:12,Yuan:14,Yuan:15}  and observational (see references in
\S\ref{subsubsec:hotwind}) aspects. In other words, whenever we observe jet, wind must also exist.
So the formation of X-ray cavity shown in Fig. \ref{fig:xsurf} suggests that it is worthwhile
to investigate the possibility that the cavities usually observed are formed by wind. In fact,
winds launched from hot accretion flow have been used  to successfully explain the formation of
the Fermi bubbles observed in the Galaxy \citep{Mou:14, Mou:15}. Compared with other models of
the Fermi bubbles \citep[e.g.,][]{Guo:12}, this ``wind'' model has the following advantages: 1)
the main parameters of the model such as mass flux and velocity of wind  are well constrained
by the small-scale MHD numerical simulation  \citep{Yuan:15}, thus has much less freedom; 2)
the model can successfully explain some observations of the Fermi bubbles that are hard to be
explained by other models (see \citealt{Mou:15} for details).

\section{Summary and Conclusions}

In this paper, by performing two-dimensional high-resolution hydrodynamical numerical simulations,
we have investigated the effects of AGN feedback in the evolution of its host galaxy.  The galaxy
is an isolated elliptical galaxy and we assume in this work that the specific angular momentum
of the gas is low. Physical processes like star formation, Type Ia and Type II supernovae are
taken into account. The inner boundary of the simulation is chosen so that the Bondi radius is
resolved, which is crucial for the precise determination of the mass accretion rate of the AGN.
According to the theory of black hole accretion, black hole accretion has two intrinsically
different modes, cold and hot ones. They have quite different radiation and wind outputs and thus
naturally correspond to two different feedback modes. They have been carefully discriminated
and taken into account in the present work. We consider the feedback effects by both radiation
and wind in each mode. The two feedback modes have quite diverse names in literature, e.g.,
the quasar (or radiative) and radio (or kinetic or maintenance) mode.    Our present work
indicates that these names are not only diverse, but sometimes also misleading. For example,
in terms of regulating the accretion rate of the black hole and star formation in the galaxy,
feedback by wind is always much more important that radiation. So we suggest to simply follow the
names of black hole accretion mode and call them ``cold feedback mode'' and ``hot feedback mode''.

The most important distinctive feature of the present work is that we adopt the most updated
AGN physics.  This is especially the case for the radiation and wind from hot accretion flows
for which many progresses have been made in recent years but have not been taken into account in
most AGN feedback works \citep[see a recent review by][]{Yuan:14}. It also includes more precise
descriptions of wind in the cold mode which are obtained from recent new observations. These
most updated AGN physics are summarized in the present paper. For the feedback effects,
we have investigated  the light curve of the AGN, the black hole growth, star formation,
the AGN duty-cycle, and the surface brightness of the galaxy. We have compared these results
with previous works which have very similar model framework with our work but have different
AGN physics. Significantly different results have been found in almost every aspect mentioned
above. This indicates the crucial importance of having correct AGN physics. The  main results
obtained in this paper are summarized below.

\begin{itemize}

\item We have compared the energy and momentum fluxes of wind and radiation in the two feedback
modes (Fig. \ref{fig:windradcomp}). Roughly speaking, in both modes, the power of radiation
is larger than  wind while the momentum flux is on the opposite. However, the magnitude of
energy or momentum fluxes is not the only factor to determine which component, radiation or
wind, is more important in the feedback.  This is because the cross section of photon-particle
interaction is orders of magnitude smaller than that of particle-particle interaction. For
typical parameters of our problem, we find that wind can deposit its momentum within a very small
``typical length scale of wind feedback'', $l_{\rm wind}\sim 0.5$ pc (Eq. \ref{windlength});
while the ``typical length scale of radiation feedback'' is much larger, $l_{\rm rad}\sim 10$ kpc
(Eq. \ref{radlength}). Consequently, in our model the accretion rate and the mass growth of black
hole are mainly suppressed by wind rather than radiation (compare the zoom in plots of windFB,
fullFB, and radFB in Fig. \ref{fig:ldot};  and windFB, fullFB, and radFB in Fig. \ref{fig:bhmass}).
Such a result is in good agreement with \citet{Gan:14} and \citet{Ciotti:17}. But we note there
are two caveats here. One is that in the present paper we  consider an isolated galaxy without
external gas supply. Another one is that we do not consider dust. If we consider these two
factors, the radiative feedback should become relatively more important.

\item One characteristic consequence when the AGN feedback is included in the galaxy evolution
model is that the AGN activity becomes strongly variable (Fig. \ref{fig:ldot}). The reason is
because of the interaction between wind \& radiation and ISM (\S\ref{overallscenario}). Both
radiation or wind can cause the variability of AGN but their mechanisms are different. Wind is by
momentum interaction while radiation is by radiative heating. Because the typical length scale
of wind feedback is much shorter than that of radiation, the accretion rate of AGN is reduced
to a much lower time-averaged value in the windFB model  than in the radFB model (right panel
of Fig. \ref{fig:ldot}). But radiation also plays an important role, maybe by coupling with
the wind feedback. Comparison with previous works which have different AGN physics indicates
that stronger wind in the AGN physics results in a time-averaged weaker AGN (compare the zoom
in plot of fullFB and B05v in Fig. \ref{fig:ldot}).

\item The typical lifetime of the AGN obtained by our simulation is $\sim 10^5$ yr,  fully
consistent with the most recent observations  (Fig. \ref{fig:agnlifetime}). As a comparison,
previous works which have different AGN physics, such as \citet{Gan:14}, obtain a much longer
AGN lifetime.

\item Both radiation and wind can suppress the star formation in the galaxy. In our model,
radiation can affect the region of ``unity optical depth'',  $\sim 1$ kpc, while wind can
affect a much larger region, $\la 20$ kpc (the left plot of Fig. \ref{fig:newst}). Not only
the affected scale is different, the wind can also suppress the star formation much more
strongly than radiation (the left plot of Fig. \ref{fig:newst}).  Beyond $20~{\rm kpc}$,
star formation is enhanced, because of the accumulation of gas pushed out by wind.  Overall,
we find that the time-integrated total effect of AGN feedback on star formation is negative
(right panel of Fig. \ref{fig:newst}). But we would like to emphasize that whether the feedback
can suppress or enhance star formation not only depends on the spatial location but also on
time. The spatially-integrated specific star formation rate (sSFR) as a function of time shows
two important features. One is that depending on the evolution time, sSFR can be enhanced or
suppressed compared to the model without AGN feedback. The second one is that radiation also
plays a very important role in affecting star formation (compare windFB and fullFB models in
Fig. \ref{fig:sSFR}).

\item  When all the feedback mechanisms have been considered, we find that the AGNs spend over
80\% of their time with Eddington ratio below $2\times 10^{-4}$, i.e., in the very low-luminosity
regime (the left panel of Fig. \ref{fig:dutycycle}). This suggests the importance of considering
the hot mode feedback in the galaxy evolution. We have also compared the simulated percentage
of the last 2 Gyrs spent above a given Eddington ratio with observations. We find that the
AGN has very little of its time being in the range of $L_{\rm BH}/L_{\rm Edd}\ga 1\%$,  consistent with
observations (right panel of Fig. \ref{fig:dutycycle}). We have also calculated the percentage of
the total energy emitted above a given Eddington ratio. For our fullFB model, it is only $6$\%
at Eddington ratio above 0.02. This value is not consistent with what we  believe. An important
reason is that we assume an isolated galaxy and have not taken into account the external gas
supply in our simulations.

\item We have calculated the X-ray luminosity of the hot gas in the galaxy in the 0.5-2 keV
band and compared the result with observations. Both models with and without AGN feedback are
consistent with observations given their large error bars (Fig. \ref{fig:xlumcomp}).  But once
we include external gas supply in the future, the noFB model may overpredict the observed value.

\item The X-ray surface brightness  for the model fullFB during an AGN outburst is calculated
(Fig. \ref{fig:xsurf}). An X-ray cavity surrounding the AGN is evident, which is formed by
the interaction between wind and ISM. It looks very similar to the  X-ray cavities observed in
galaxy clusters \citep[e.g.,][]{Fabian:12}.  Usually we think these X-ray cavities are formed
by jets. So this result suggests us to consider the possibility that they are formed by wind.


\end{itemize}

\section*{Acknowledgements}

We thank the referee for his/her careful reading and constructive comments, which have
significantly improved our paper. We are grateful to Jerry Ostriker and Luca Ciotti  who kindly
sent us the early version of the code in which many basic physics involved in the modeling
are presented. This work is supported in part by the National Key Research and Development
Program of China (grants 2016YFA0400702 and 2016YFA0400704),  the Natural Science Foundation
of China (grants 11573051, 11633006, 11650110427, 11661161012, 11303008, 11473002),  and the
Key Research Program of Frontier Sciences of CAS (grants  QYZDJSSW- SYS008 and QYZDB-SSW-SYS033
). This work has made use of the High Performance Computing Resource in the Core Facility for
Advanced Research Computing at Shanghai Astronomical Observatory.

\bibliographystyle{aasjournal}
\bibliography{AGNHot_arXiv}

\end{document}